\documentclass[aps,pra,superscriptaddress,twocolumn,showpacs,floatfix,10pt]{revtex4-1}

\usepackage{amsmath}
\usepackage{amssymb}
\usepackage{bm}
\usepackage{graphicx}
\usepackage{appendix}
\usepackage{longtable}

\begin{document}


\title{Nonlinear optical response of a two-dimensional quantum dot supercrystal:\\
Emerging multistability, periodic/aperiodic self-oscillations, and hyperchaos}

\author{Pablo \'Alvarez Zapatero}
\affiliation{GISC, Departamento de F\'{\i}sica de Materiales,
Universidad Complutense, E-28040 Madrid, Spain}

\author{Ramil F. Malikov}
\affiliation{Akmullah State Pedagogical University of Bashkortostan, 450000 Ufa, Russia}

\author{Igor V. Ryzhov}
\affiliation{Hertsen State Pedagogical University, 191186 St. Petersburg, Russia}

\author{Andrey V. Malyshev}
\email[Corresponding author:]{a.malyshev@fis.ucm.es}
\affiliation{GISC, Departamento de F\'{\i}sica de Materiales,
Universidad Complutense, E-28040 Madrid, Spain} \affiliation{Ioffe
Physical-Technical Institute, 26 Politechnicheskaya str., 194021
St. Petersburg, Russia}

\author{Victor\ A.\ Malyshev}
\email[Corresponding author:]{v.malyshev@rug.nl}
\affiliation{Zernike Institute for Advanced Materials, University of Groningen, Nijenborgh 4, 9747
AG Groningen, The Netherlands}

\date{\today}

\begin{abstract}
We study theoretically the nonlinear optical response of a two-dimensional semiconductor quantum dot supercrystal under a resonant continuous wave excitation. A single quantum dot is modeled as a three-level ladder-like system with the ground, one-exciton, and bi-exction states. We propose an exact linear parametric method of solving the nonlinear steady-state problem. It is demonstrate that the system may exhibit multistability, periodic and aperiodic self-oscillations, and hyperchaotic behavior, depending on the system's parameters and frequency of excitation. The effects originate from the retarded dipole-dipole interaction of quantum dots. The latter provides a positive feedback which, in combination with the nonlinearity of SQDs, leads to an exotic nonlinear dynamics of the system indicated above. We discuss relevance of the underlined effects for nanosized all-optical devices. In particular, a quantum dot supercrystal may serve as a nanosized all-optical switch, a tunable generator of trains of THz pulses (in self-oscillating regime), as well as a noise generator (in chaotic regime) at the nanoscale. We show also that the supercrystal can operate as a bistable mirror. All this suggests various nanophotonic applications of such type of materials. 

\end{abstract}

\maketitle

\section{Introduction}
\label{Intro}
\begin{figure}[ht!]
\begin{center}
\includegraphics[width=\columnwidth]{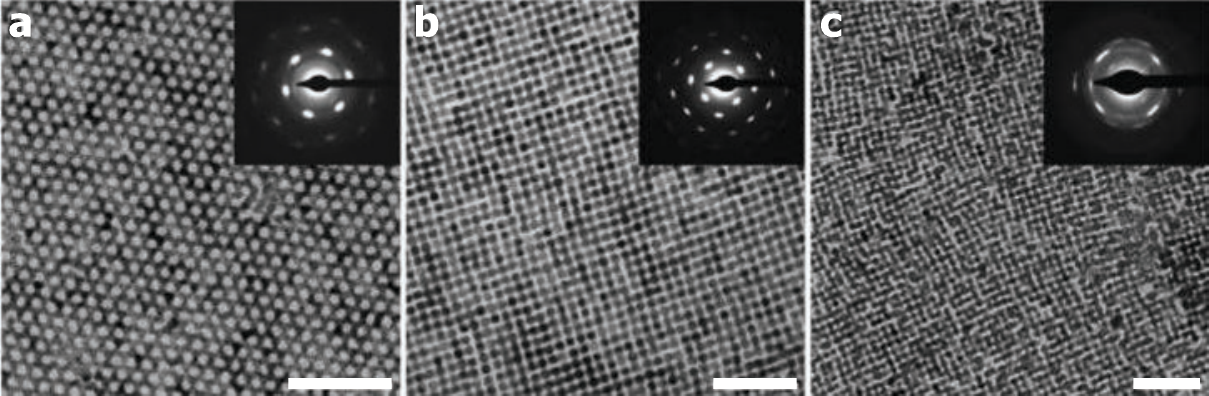}
\caption{\label{fig:Supercrystal} (a) Ultrathin PbSe rocksalt semiconductor sheet with a honeycomb nanostructuring. (b) Ultrathin PbSe rocksalt semiconductor sheet with a square nanostructuring. (c) Ultrathin CdSe semiconductor with a compressed zincblende atomic structure and slightly distorted square nanostructuring. White scale bars represent 50 nm. Insets show the electro-diffractograms in the [111] (a) and [100] (b,c) projections. The picture is taken from Ref.~\onlinecite{Evers2013} (for more details, see Ref.~\onlinecite{Evers2013}).}
\end{center}
\end{figure}

In the last decade, the so-called metamaterials, a class of new materials not existing in nature, received a great deal of attention (see for recent reviews Refs.~\onlinecite{Zheludev2010,Zheludev2012,Soukoulis2010,Liu2011,Alu2016}). Super-crystals comprising regularly spaced quantum emitters represent one of examples of metamaterials with tunable optical properties which can be controlled by the geometry and chemical composition of components.~\cite{Baimuratov2013} Modern nanotechnology has in its disposal a variety of methods to fabricate such systems.~\cite{Evers2013,Boneschanscher2014,Baranov2015,Ushakova2016,Liu2017} In Fig.~\ref{fig:Supercrystal}, a few examples of  ultrathin sheets of regularly spaced semiconductor nanocrystals grown by the method of oriented attachment (see for details Ref.~\onlinecite{Evers2013}) are present.

As is well known, a thin layer of two-level emitters (atoms, molecules, J-aggregates), the thickness $L$ of which is much smaller than the radiation wavelength $\lambda$ in the layer, may act as an all-optical bistable element.~\cite{Bowden1986a,Bowden1986b,Zakharov1988,Basharov1988,Benedict1990,Benedict1991,Oraevsky1994,Malyshev2000,Glaeske2000,Klugkist2007, Malikov2017a} For bistability to occur, two factors are required: nonlinearity of the material and a positive feedback. Interplay of these two factors leads to a situation when the system has two stable states; switching between them is governed again by an external optical signal. The nonlinearity of the layer is ensured by the fact that two-level emitters are nonlinear systems. The positive feedback originates from the secondary field, which is generated by the emitters themselves; this is the so-called “intrinsic feedback”, i.e. here, a cavity (external feedback) is not required.

A two-dimensional (2D) semiconductor quantum dot (SQD) supercrystal represents a limiting case of a thin layer. In this paper, we conduct a theoretical study of the nonlinear optical response of such a system. A single SQD dot is considered as a point-like system with three consecutive levels of the ground, one-exciton, and bi-exction states (corresponding to the so-called ladder or $\Theta$ level scheme). Due to the high density of SQDs and high oscillator strengths of the SQD's transitions, the total (retarded) dipole-dipole SQD-SQD interactions have to be taken into account, which is finally done in the mean-field approximation for the point-like dipoles in a homogeneous host for simplicity. The real part of the dipole-dipole interaction results in the dynamic shift of the SQD's energy levels, whereas the imaginary part describes the collective radiative decay of SQDs, both depending on the population differences between the levels. These two effects are crucial for the nonlinear dynamics of the SQD supercrystal. As a result, in addition to bistability, analogous to that for a thin layer of two-level emitters, we found multistability, periodic and aperiodic self-oscillation, and chaotic regimes in the optical response of the SQD supercrystal.~\cite{footnote} To uncover the character of the instabilities, we analyze the Lyapunov exponents of different branches of the steady-state characteristics, as well as the phase space map of the full field and Fourier spectrum of the latter. An important achievement of our study is the invention of a method of finding the exact solution of the nonlinear steady-state multilevel Maxwell-Bloch equations for a layer-like system. To the best of our knowledge, a detailed study of the SQD supercrystal optical response has not been performed so far.~\cite{Malikov2017b} 

The arrangement of the paper is as follows. In the next section, we describe the model of a 2D supercrystal comprised of SQDs and the mathematical formalism to treat its optical response. We use for that the one-particle density matrix formalism within the rotating wave approximation (RWA), where the total retarded dipole-dipole interactions between point-like SQDs are taken into account. In Sec.\ref{Mean-field approximation}, the general formalism is simplified making use of the mean-field approximation, and the mean-field parameters (the collective energy level shift and radiation damping) are calculated. In Sec.~\ref{Numerics}, we present the results of numerical calculations of the supersrystal optical response for two conditions of excitation: (i) the external field is tuned into the one-exciton transition and (ii) it is in resonance with the coherent two-photon transition (with the simultaneous absorption of two photons). The results are discussed in Sec.~\ref{Discussion}. In Sec.~\ref{Summary}, we conclude and discuss the possibility to observe the effects found in the paper in real-world nanosystems.

\section{Model and theoretical background}
\label{Model}
\begin{figure}
\begin{center}
\includegraphics[width=0.6\columnwidth]{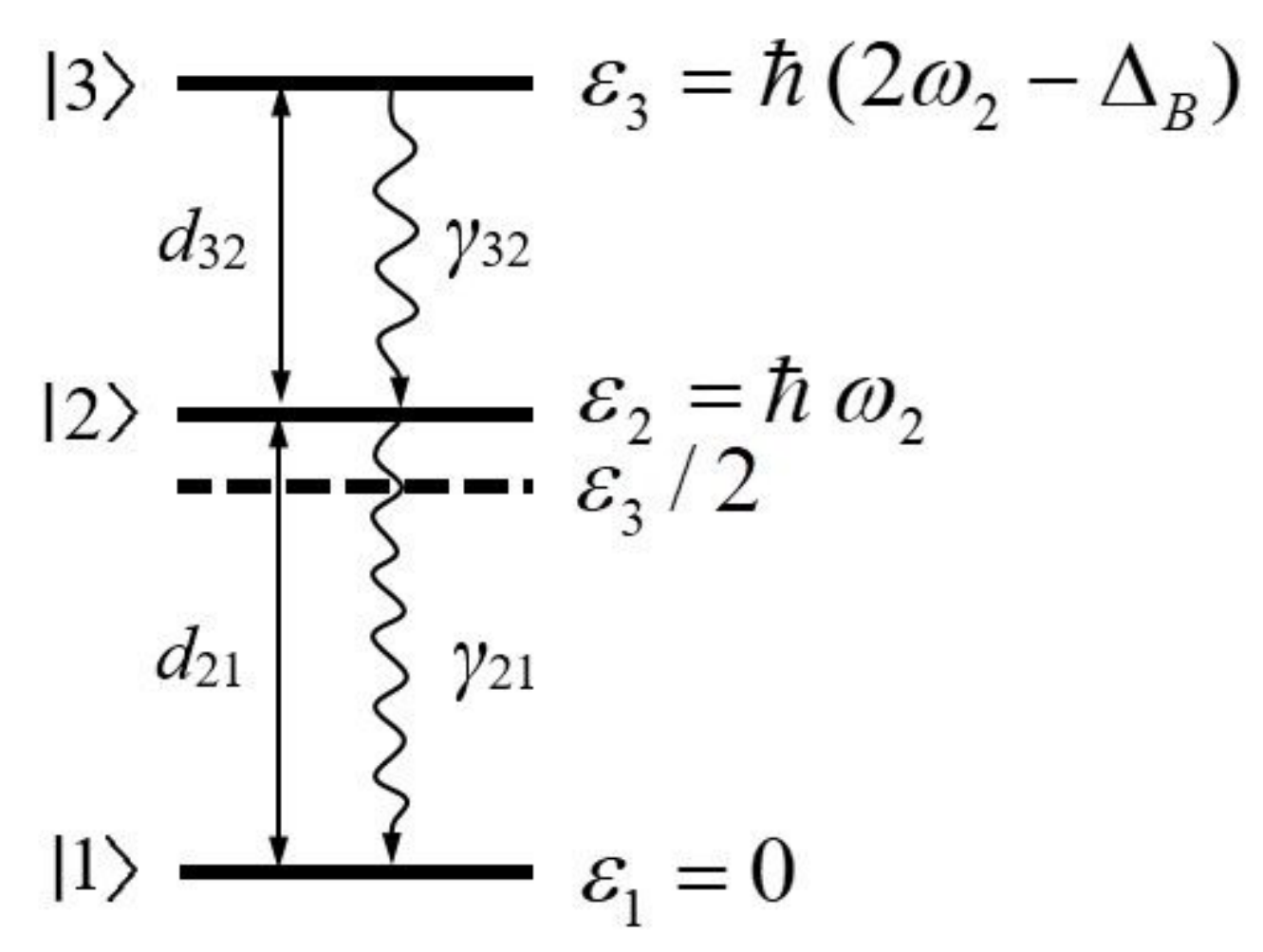}
\caption{\label{fig:Schematics} Energy level diagram of a ladder-type three-level SQD: $|1 \rangle$, $|2 \rangle$, and $|3 \rangle$ are the ground, one-exciton, and bi-exciton  states, respectively. The energies of corresponding states are $\varepsilon_1 =0$, $\varepsilon_2 =\hbar \omega_2$ and $\varepsilon_3 = \hbar(2\omega_2 -\Delta_B)$. Here, $\hbar\Delta_B$ is the bi-exciton  binding energy. Allowed transitions with the corresponding transition dipole moments $\bf{d}_{21}$ and $\bf{d}_{32}$ are indicated by solid arrows. Wigging arrows denote the allowed spontaneous transitions with rates $\gamma_{32}$ and $\gamma_{21}$. The dashed horizontal line shows the location of the coherent two-photon resonance (with simultaneous absorption of two photons).}
\end{center}
\end{figure}

We consider a 2D supercrystal comprising identical semiconductor quantum dots (SQDs).
The optical excitations in an SQD are confined excitons. In such a system, the degeneracy of the one-exciton state is lifted due to the anisotropic electron-hole exchange, leading to two split linearly polarized one-exciton states (see, e.g., Refs.~\onlinecite{Stufler2006,Jundt2008,Gerardot2009}). In this case, the ground state is coupled to the bi-exciton state via the linearly polarized one-exciton transitions. By choosing the appropriate polarization of the applied field, i.e., selecting one of the single-exciton states, the system effectively acquires a three-level ladder-like structure with a ground state $|1\rangle$, one exciton state $|2\rangle$, and bi-exciton state $|3\rangle$ with corresponding energies $\varepsilon_1 = 0$, $\varepsilon_2 = \hbar\omega_{2}$, and $\varepsilon_3 = \hbar\omega_3 = \hbar(2\omega_2 - \Delta_B)$, where $\hbar\Delta_B$ is bi-exciton binding energy (see Fig.~\ref{fig:Schematics}). Within this model, the allowed transitions, induced by the external field, are $|1\rangle \leftrightarrow |2\rangle$ and $|2\rangle \leftrightarrow |3\rangle$, which are characterized by the transition dipole moments $\bf{d}_{21}$ and $\bf{d}_{32}$, respectively. For the sake of simplicity, we assume that they are real. The states $|3\rangle$ and $|2\rangle$ spontaneously decay with rates $\gamma_{32}$ and $\gamma_{21}$, accordingly. Note that the bi-exciton  state $|3\rangle$, having no allowed transition dipole moment from the ground state $|1\rangle$, can be populated either via consecutive $|1\rangle \rightarrow |2\rangle \rightarrow |3\rangle$ transitions or via the simultaneous absorption of two photons of frequency $\omega_3/2$. In what follows, we will consider both options.

The optical dynamics of SQDs is described by means of the Lindblad quantum master equation for the density operator $\rho(t)$,~\cite{Lindblad1976,Blum2012} which in the rotating frame (with the frequency of the external field $\omega_0$) reads
\begin{widetext}
\begin{subequations}
\label{MasterEqAndHamiltonian}
\begin{equation}
\dot{\rho}(t) = -\frac{i}{\hbar} \left[{H^{RWA}}(t),\rho(t)\right ] + {\cal L}\{\rho(t)\}~,
\label{LindbladEq}
\end{equation}
\begin{equation}
H^{RWA}(t) =  \hbar \sum_{\bf n} \left( \Delta_{21}\sigma^{\bf n}_{22} + \Delta_{31}\sigma^{\bf n}_{33} \right)
        - i\hbar \sum_{\bf n} \left[ {\Omega}^{\bf n}_{21}(t) \sigma^{\bf n}_{21} + \Omega^{\bf n}_{32}(t) \sigma^{\bf n}_{32} \right] + h.c.
\label{HamiltonianRWA}
\end{equation}
\begin{equation}
{\cal L}\{\rho(t)\} =
           \frac{\gamma_{21}}{2} \sum_{\bf n} \left( \left[ \sigma^{\bf n}_{12} \rho(t),
           \sigma^{\bf n}_{21}\right]  + \left[ \sigma^{\bf n}_{12},
           \rho(t)\, \sigma^{\bf n}_{21}\right]\right)
           + \frac{\gamma_{32}}{2} \sum_{\bf n} \left( \left[
           \sigma^{\bf n}_{23}\rho(t),\sigma^{\bf n}_{32}\right]
           + \left[ \sigma^{\bf n}_{23},\rho(t) \, \sigma^{\bf n}_{32}\right]\right)
\label{LindbladOperator}
\end{equation}
\begin{equation}
\sigma^{\bf n}_{ij} = |{\bf n}i\rangle \langle j{\bf n}|~, \quad i,j = 1,2,3~.
\end{equation}
\end{subequations}
\end{widetext}
In Eq.~(\ref{LindbladEq}), $\hbar$ is the reduced Plank constant, $H^{RWA}$ is the SQD Hamiltonian in the RWA, $[A,B]$ denotes the commutator, ${\cal L}$ is the Lindblad relaxation operator, given by Eq.~(\ref{LindbladOperator}).~\cite{Lindblad1976,Blum2012}
In Eq.~(\ref{HamiltonianRWA}), $ \Delta_{21} = \omega_2 - \omega_0$ and $\Delta_{31} = \omega_3 - 2\omega_0$ are the energies of states $|2 \rangle$ and $|3 \rangle$ in the rotating frame, accordingly. Alternatively, these quantities may be interpreted as
the detunings away from the one-photon resonance and the coherent two-photon resonance, respectively. $\Omega^{\bf n}_{21}(t) = \bf{d}_{21} \bf{E_{n}}(t)/\hbar$ and $\Omega^{\bf n}_{32}(t) = \bf{d}_{32}\bf{E_n}(t)/\hbar$, where $\bf{E_n}(t)$ is the slowly-varying amplitude of the total field driving the optical transitions in the ${\bf n}$-th SQD, $\bm{\mathcal E}_{\bf n}(t) = {\bf E_n}(t)\exp(-i\omega_0 t) + c.c.$. The latter is the sum of the applied field $\bm{\mathcal E}^0_{\bf n}(t) = {\bf E}^0_{\bf n}(t)\exp(-i\omega_0 t) + c.c.$ and the field produced by all others SQDs in place of the ${\bf n}$-th SQD,
$\bm{\mathcal E}^{loc}_{\bf n}(t) = \sum_{\bf m} \bm{\mathcal E}_{\bf mn}(t) = \sum_{\bf m} {\bf E_{mn}}(t)\exp(-i\omega_0 t) + c.c.$, where the amplitude ${\bf E_{mn}}(t)$ is given by (see, e.g., Refs.~\onlinecite{Zaitsev1983,Benedict1996})~\footnote{\label{note1}Strictly speaking, this field should be the field acting {\it inside} the SQD, the latter differs from the field acting {\it on} the SQD by a screening factor which depends on the system geometry and material parameters. In the simplest case of a spherical dot in a homogeneous environment this factor can be obtained analytically (see, e. g., Ref.~\onlinecite{Malyshev2011}). A realistic SQD array is a considerably more complicated system involving a non-homogeneous host, at least three different materials, and a number of geometrical parameters. We believe that explicit calculation of the screening factors in this case would introduce unnecessary level of detailization and obscure further analysis. Therefore, for the sake of simplicity, we consider a SQD as a point-like system in a homogeneous host; all the fields entering the Lindblad equations should be interpreted as those rescaled by appropriate screening factors.}
\begin{widetext}
\begin{eqnarray}
\label{Emn}
{\bf E_{mn}}(t) = \left\{\left[ \frac{3}{r^3_{\bf mn}} - \frac{3ik_0}{r^2_{\bf mn}}  - \frac{k_0^2}{r_{\bf mn}}\right]({\bf d_{21}u_{mn}}){\bf u_{mn}}
- \left[ \frac{1}{r^3_{\bf mn}} - \frac{ik_0}{r^2_{\bf mn}} - \frac{k_0^2}{r_{\bf mn}}\right]{\bf d_{21}}  \right\}\rho^{\bf m}_{21}(t^\prime)e^{ik_0r_{\bf mn}} \nonumber
\\
+ \left\{\left[ \frac{3}{r^3_{\bf mn}} - \frac{3ik_0}{r^2_{\bf mn}}  - \frac{k_0^2}{r_{\bf mn}}\right]({\bf d_{32}u_{mn}}){\bf u_{mn}}
- \left[ \frac{1}{r^3_{\bf mn}} - \frac{ik_0}{r^2_{\bf mn}} - \frac{k_0^2}{r_{\bf mn}}\right]{\bf d}_{32}  \right\}\rho^{\bf m}_{32}(t^\prime) e^{ik_0r_{\bf mn}}~,
\end{eqnarray}
\end{widetext}
where $r_{\bf mn}$ is the distance between sites ${\bf m}$ and ${\bf n}$, $k_0 = \omega_0/c$ ($c$ is the speed of light in vacuum), ${\bf u_{mn}} = {\bf r_{mn}}/r_{\bf mn}$ is the unit vector along ${\bf r_{mn}}$, and $t^\prime = t - r_{\bf mn}/c$. Equation~(\ref{Emn}) represents the field (amplitude) produced by an oscillating dipole ${\bf d_{21}}R^{\bf m}_{21}(t^\prime) + {\bf{d_{32}}}R^{\bf m}_{32}(t^\prime)$ situated at a point ${\bf r_m}$ in another point ${\bf r_n}$ at an instant $t$, accounting for retardation: $t - t^\prime = r_{\bf mn}/c$. Having Eq.~(\ref{Emn}), the fields $\Omega^{\bf n}_{\alpha\beta}(t)$ ($\alpha\beta = 21,32$) can be written in the form
\begin{widetext}
\begin{subequations}
\begin{equation}
\label{Omega21n}
\Omega^{\bf n}_{\alpha\beta}(t) = \Omega^{0{\bf n}}_{\alpha\beta}(t)
    + \sum_{\bf m} (\gamma_{\alpha\beta}^{\bf mn} + i\Delta_{\alpha\beta}^{\bf mn}) \rho_{\alpha\beta}^{\bf m}(t^\prime)~,
\end{equation}
\begin{equation}
\label{gamma21n}
\gamma_{\alpha\beta}^{\bf mn} = \frac{3\gamma_{\alpha\beta}}{4(k_0a)^3} \left\{ \left[(k_0a)^2 \frac{\kappa_{\alpha\beta}^{\mathbf mn}}{|{\bf m-n}|}
- \frac{\chi_{\alpha\beta}^{\mathbf mn}}{|{\bf m-n}|^3} \right] \sin(k_0a|{\bf m-n}|) + k_0a \frac{\chi_{\alpha\beta}^{\mathbf mn}}{|{\bf m-n}|^2}\cos(k_0a|{\bf m-n}|) \right\}
\end{equation}
\begin{equation}
\label{Delta21n}
\Delta_{\alpha\beta}^{\bf mn} = \frac{3\gamma_{\alpha\beta}}{4(k_0a)^3} \left\{ \left[ \frac{\chi_{\alpha\beta}^{\mathbf mn}}{|{\bf m-n}|^3}
- (k_0a)^2 \frac{\kappa_{\alpha\beta}^{\mathbf mn}}{|{\bf m-n}|} \right] \cos(k_0a|{\bf m-n}|) + k_0a \frac{\chi_{\alpha\beta}^{\mathbf mn}}{|{\bf m-n}|^2}\sin(k_0a|{\bf m-n}|) \right\}
\end{equation}
\begin{equation}
\label{kappa21mn}
\kappa_{\alpha\beta}^{\mathbf mn} = 1 - ({\bf e}_{\alpha\beta}{\bf u_{mn}})^2~,
    \quad \chi_{\alpha\beta}^{\mathbf mn} = 1 - 3({\bf e}_{\alpha\beta}{\bf u_{mn}})^2~.
\end{equation}
\end{subequations}
\end{widetext}
We used in Eqs.~(\ref{gamma21n}) and~(\ref{Delta21n}) the expression $\gamma_{\alpha\beta} =
4|{\bf d}_{\alpha\beta}|^2 \omega_{\alpha\beta}^3/(3\hbar c^3)$.
In Eq.~(\ref{kappa21mn}), ${\bf e}_{\alpha\beta} = {\bf d}_{\alpha\beta}/d_{\alpha\beta}$ is the unit vector along ${\bf d}_{\alpha\beta}$.
The matrices $\gamma_{\alpha\beta}^{\bf mn}$ and $\Delta_{\alpha\beta}^{\bf mn}$ represent the real and imaginary parts of the retarded dipole-dipole interaction of $\bf n$-th and $\bf m$-th SQDs.

Equation~(\ref{LindbladEq}), written in the site basis $|{\bf n}i \rangle$ ($i = 1,2,3$), reads
\begin{widetext}
\begin{subequations}
\label{allrhon}
\begin{equation}
\label{rho11n}
\dot{\rho}^{\bf n}_{11} = \gamma_{21} \rho^{\bf n}_{22} + \Omega^{\bf n}_{21} {\rho^{\bf n *}_{21}} + {\Omega^{\bf n *}_{21}} \rho^{\bf n}_{21}~,
\end{equation}
\begin{equation}
\label{rho22n}
\dot{\rho}^{\bf n}_{22} =  -\gamma_{21} \rho^{\bf n}_{22} + \gamma_{32} \rho^{\bf n}_{33} - \Omega^{\bf n}_{21} \rho^{\bf n*}_{21}
    - \Omega^{\bf n*}_{21} \rho^{\bf n}_{21} + \Omega^{\bf n}_{32} \rho^{\bf n*}_{32} + \Omega^{\bf n*}_{32} \rho^{\bf n}_{32}~,
\end{equation}
\begin{equation}
\label{rho33n}
\dot{\rho}^{\bf n}_{33} = -\gamma_{32} \rho^{\bf n}_{33} - \Omega^{\bf n}_{32} \rho^{\bf n *}_{32} - \Omega^{\bf n*}_{32} \rho_{32}~,
\end{equation}
\begin{equation}
\label{R21n}
\dot{\rho}^{\bf n}_{21} = - \left( i\Delta_{21} + \frac{1}{2}\gamma_{21} \right) \rho^{\bf n}_{21} + \Omega^{\bf n}_{21}(\rho^{\bf n}_{22}
    - \rho^{\bf n}_{11}) + \Omega^{\bf n*}_{32} \rho_{31}~,
\end{equation}
\begin{equation}
\label{R32n}
\dot{\rho}^{\bf n}_{32} = - \left[ i\Delta_{32} + \frac{1}{2} (\gamma_{32} + \gamma_{21}) \right] \rho^{\bf n}_{32}
    +\Omega^{\bf n}_{32}(\rho^{\bf n}_{33} - \rho^{\bf n}_{22}) - \Omega^{\bf n *}_{21} \rho^{\bf n}_{31}~,
\end{equation}
\begin{equation}
\label{R31n}
\dot{\rho}^{\bf n}_{31} = - \left( i\Delta_{31} + \frac{1}{2}\gamma_{32} \right) \rho^{\bf n}_{31} - \Omega^{\bf n}_{32} \rho^{\bf n}_{21}
    + \Omega^{\bf n}_{21} \rho^{\bf n}_{32}~,
\end{equation}
\end{subequations}
\end{widetext}
where $\Omega^{\bf n}_{21}$ and $\Omega^{\bf n}_{32}$ are given by Eqs.~(\ref{Omega21n}) -~(\ref{kappa21mn}). The time dependence of all relevant quantities is dropped here.

It is worth to noting that Eqs.~(\ref{rho11n}) -~(\ref{R31n}) represent a set of equations for the one-particle density matrix, where the quantum correlations of the dipole operators of different SQDs are neglected that implies that $\langle \hat{d}_{\bf n} \hat{d}_{\bf m}\rangle = \langle \hat{d}_{\bf n}\rangle\langle \hat{d}_{\bf m}\rangle$, where $\langle ... \rangle$ denotes the quantum mechanical average. A prove of this assumption is a stand-alone problem to be solved, that is is beyond the scope of the present paper.

\section{Mean-field approximation}
\label{Mean-field approximation}
The set of Eqs.~(\ref{rho11n}) - (\ref{R31n}) allows one to study the optical response of a SQD monolayer, without any limitation to the layer's size, lattice geometry, and the spatial profile of the external field amplitude ${\bf E}^{\bf n}_0$. Here, we restrict our consideration to a spatially homogeneous case, when all relevant quantities entering Eqs.~(\ref{rho11n}) - (\ref{R31n}) do not depend on the SQD's position $\bf n$. In fact, this approximation is equivalent to taking into account the Lorentz local field correction to the field acting on an emitter, which has been widely used when analysing the optical response of dense media, both linear~\cite{BornAndWolf,Friedberg1973} and nonlinear.~\cite{Hopf1984,Ben-Aryeh1986,Benedict1990,Malyshev1997a,Malyshev1997b,Malikov2017a} This approximation intuitively seems to be appropriate for an infinite layer, however, for a finite sample, its validity should be examined. Nevertheless, as will be shown below, even this simplest model already brings to the layers's optical response a variety of fascinating effects. Not to complicate the computational work, we consider a simple square lattice of SQDs.

Thus, we neglect the spatial dependence of all functions in Eqs.~(\ref{rho11n}) - (\ref{R31n}). Additionally, we assume for the sake of simplicity that the transition dipoles ${\bf d}_{21}$ and ${\bf d}_{32}$ are parallel to each other, ${\bf d}_{32} = \mu {\bf d}_{21} \equiv \mu{\bf d}$ (not a principal limitation). Accordingly, $\gamma_{32} = \mu^2 \gamma_{21} \equiv \mu^2 \gamma$ and $\Omega_{32} = \mu \Omega_{21} \equiv \mu \Omega$. Then the system of equations~(\ref{rho11n}) - (\ref{R31n}) takes the form~\cite{footnote2}
\begin{widetext}
\begin{subequations}
\label{allrho}
\begin{equation}
\label{rho11}
\dot{\rho}_{11} = \rho_{22} + \Omega \rho_{21}^* + \Omega^* \rho_{21}~,
\end{equation}
\begin{equation}
\label{rho22}
\dot{\rho}_{22} =  - \rho_{22} + \mu^2 \rho_{33} - \Omega\rho_{21}^* - \Omega^* \rho_{21} + \mu (\Omega\rho_{32}^* + \Omega^*\rho_{32})~,
\end{equation}
\begin{equation}
\label{rho33}
\dot{\rho}_{33} = -\mu^2 \rho_{33} - \mu (\Omega \rho_{32}^* + \Omega^* \rho_{32})~,
\end{equation}
\begin{equation}
\label{R21}
\dot{\rho}_{21} = - \left( i\Delta_{21} + \frac{1}{2} \right) \rho_{21} + \Omega(\rho_{22} - \rho_{11}) + \mu \Omega^* \rho_{31}~,
\end{equation}
\begin{equation}
\label{R32}
\dot{\rho}_{32} = - \left[ i\Delta_{32} + \frac{1}{2} (1 + \mu^2) \right] \rho_{32} + \mu \Omega(\rho_{33} - \rho_{22}) - \Omega^* \rho_{31}~,
\end{equation}
\begin{equation}
\label{R31}
\dot{\rho}_{31} = - \left( i\Delta_{31} + \frac{1}{2}\mu^2 \right) \rho_{31} - \mu \Omega \rho_{21} + \Omega \rho_{32}~,
\end{equation}
\begin{equation}
\label{Omega21}
\Omega = \Omega_0 + (\gamma_R + i\Delta_L) (\rho_{21} + \mu \rho_{32})~,
\end{equation}
\end{subequations}
\end{widetext}
where the constants $\gamma_R$ and $\Delta_L$ are given by
\clearpage
\begin{widetext}
\begin{subequations}
\label{gammaRDeltaL}
\begin{equation}
\label{gammaR}
\gamma_R = \sum_{\bf m(\neq n)} \gamma_{21}^{\bf mn} =
\frac{3\gamma}{4(k_0a)^3} \sum_{\bf n \neq 0} \left\{ \left[(k_0a)^2 \frac{\kappa_\mathbf n}{|{\bf n}|}
- \frac{\chi_{\mathbf n}}{|{\bf n}|^3} \right] \sin(k_0a|{\bf n}|) + k_0a \frac{\chi_{\mathbf n}}{|{\bf n}|^2}\cos(k_0a|{\bf n}|) \right\}~,
\end{equation}
\begin{equation}
\label{DeltaL}
\Delta_L = \sum_{\bf m(\neq n)} \Delta_{21}^{\bf mn} = \frac{3\gamma}{4(k_0a)^3} \sum_{\bf n \neq 0} \left\{ \left[ \frac{\chi_{\mathbf n}}{|{\bf n}|^3}
- (k_0a)^2 \frac{\kappa_\mathbf n}{|{\bf n}|} \right] \cos(k_0a|{\bf n}|) + k_0a \frac{\chi_{\mathbf n}}{|{\bf n}|^2}\sin(k_0a|{\bf n}|) \right\}~,
\end{equation}
\begin{equation}
\label{kappa}
\kappa_{\mathbf n} = 1 - ({\bf eu_n})^2~, \quad\quad \chi_{\mathbf n} = 1 - 3({\bf eu_n})^2~.
\end{equation}
\end{subequations}
\end{widetext}
Recall that the summation in Eqs.~(\ref{gammaR}) and~(\ref{DeltaL}) runs over sites ${\bf n} = (n_x,n_y)$ of a simple square lattice, where $n_x = 0, \pm 1, \pm 2, \pm 3 ...\pm N_x$, $n_y = 0, \pm 1, \pm 2, \pm 3 ...\pm N_y$, and ${\bf e} = {\bf d}/d$ is the unit vector along the transition dipole moment ${\bf d}_{21}$.

Next, we are interested in ($k_0a$)-scaling of the constants $\gamma_R$ and $\Delta_L$. First, consider a point-like system, when the lateral lattice sizes $N_xa$ and $N_ya$ are much smaller that the reduced wavelength $\lambdabar = k_0^{-1}$. Then, making the expansion of sine- and cosine-functions in Eqs.~(\ref{gammaR}) and~(\ref{DeltaL}) to the lowest order with respect to $k_0a \ll 1$, one finds
\begin{subequations}
\label{gammaRDeltaLpoint-like}
\begin{equation}
\label{gammaRpoint-like}
\gamma_R =
\frac{3\gamma}{4} \sum_{\bf n \neq 0} \kappa_{\mathbf n} = \frac{3}{8}\gamma N~,
\end{equation}
\begin{eqnarray}
\label{DeltaLpoint-like}
\Delta_L &=&
\frac{3\gamma}{4(k_0a)^3} \sum_{\bf n \neq 0} \frac{\chi_{\mathbf n}}{|{\bf n}|^3}
= -\frac{3\gamma}{2(k_0a)^3}\zeta(3/2)\beta(3/2)\nonumber \\
&\simeq& -3.39 \frac{\gamma}{(k_0a)^3} = -3.39 \gamma \left(\frac{\lambdabar}{a}\right)^3~,
\end{eqnarray}
\end{subequations}
where $N = 4 N_xN_y$ is the total number of sites in the lattice, $\zeta(x)$ is the Riman $z$-function and $\beta(x) = \sum_{n=0}^\infty (-1)^n (2n + 1)^{-x}$ is the analytical continuation of the Dirichlet series.~\cite{Glasser1972} When deriving Eq.~(\ref{gammaRpoint-like}) we used the fact that $\sum_{\bf n \neq 0} \kappa_{\bf n} = N/2$. Furthermore, the formula~(\ref{DeltaLpoint-like}) follows from Eq.~(A20) of Ref.~\onlinecite{Christiansen1998} at $\theta = \pi/2$. As is seen from Eq.~(\ref{gammaRpoint-like}), $\gamma_R$ does not depend on $k_0a$; it is determined by the total number of SQDs in the lattice and describes the collective (Dicke) radiative relaxation of SQDs as all the SQD's dipoles are in phase for a point-like system.~\cite{Dicke1954,Benedict1996} Oppositely, $\Delta_L$ shows ($k_0a$)-scaling, corresponding to the near-zone dipole-dipole interaction of a given SQD with all others.

For a large system ($N_xa, N_ya \gg \lambdabar$), one has to use Eqs.~(\ref{gammaR}) and~(\ref{DeltaL}) to calculate $\gamma_R$ and $\Delta_L$, keeping all terms when performing summation. It turns out that the sums in Eqs.~(\ref{gammaRpoint-like}) and~(\ref{DeltaLpoint-like}), which contain summands proportional to $|{\bf n}|^{-1}$, converge very slowly, leading to small oscillations of $\gamma_R$ and $\Delta_L$ with the lattice size around some average values (see Appendix A). The latter are given by
\begin{subequations}
\label{gammaRDeltaLextended}
\begin{equation}
\label{gammaRextended}
\gamma_R \simeq 4.51 \frac{\gamma}{(k_0a)^2} = 4.51 \gamma \left(\frac{\lambdabar}{a}\right)^2~.
\end{equation}
\begin{equation}
\label{DeltaLextended}
\Delta_L \simeq -3.35 \frac{\gamma}{(k_0a)^3} = -3.35 \gamma \left(\frac{\lambdabar}{a}\right)^3~,
\end{equation}
\end{subequations}
As follows from Eq.~(\ref{gammaRextended}), for a large system,  the collective radiation rate $\gamma_R$ is determined by a number of SQDs within an area on the order of $\lambdabar^2$: all SQD's dipoles are in phase there. Recall that for a linear chain of emitters, $\gamma_R \sim \lambdabar/a$.~\cite{Zaitsev1983} On the contrary, the near-zone dipole-dipole interaction $\Delta_L$ changes only slightly over that one for a point-like system [compare Eq.(\ref{DeltaLextended}) with Eq.~(\ref{DeltaLpoint-like})].

It should be noticed that irrespectively of the system size, the inequality $|\Delta_L| \gg \gamma_R$ is always fulfilled for a dense system, $\lambdabar \gg a$. We will use this relationship in our analysis of the supercrystal's optical response.

\section{Numerics}
\label{Numerics}
We performed calculations of the external field dependence of the  mean field $\Omega$ for two resonance conditions: (i) — the applied  field $\Omega_0$ is in resonance with the one-exciton transition $\omega_0 = \omega_2$ ($\Delta_{21} = 0, \Delta_{32} = - \Delta_B$) of a single emitter (conventionally called in what follows as one-photon resonance) and (ii)— it is tuned to the two-photon resonance, $\omega_0 = \omega_3/2$ ($\Delta_{21} = \Delta_B/2, \Delta_{32} = - \Delta_B/2$). In reality, however, the single emitter resonance $\Delta_{21} = 0$ is read-shifted due to the near-zone SQD-SQD interactions by $|\Delta_L|$, so that the resonance in the linear low field intensity regime is defined by the condition $\Delta_{21} = |\Delta_L|$ (see Sec.~\ref{Discussion} for detail).

The set of parameters we used in our numerical calculations has been chosen on the basis of typical optical parameters of SQDs and SQDs spacing in a supercrystal (see, e.g., Fig.~\ref{fig:Supercrystal}). More specifically, for the spontaneous decay rate $\gamma$ we used the typical for SQDs (emitting in the visible) value of $\gamma \approx 3\cdot 10^9$~s$^{-1}$, whereas for the ratio $\mu = d_{32}/d_{21}  = \sqrt{\gamma_{32}/\gamma_{21}}$, a value of $\mu = \sqrt{2/3}$. The magnitudes of $\gamma_R$ and $\Delta_L$ depend on the ratio $\lambdabar/a$. Taking $\lambdabar \sim 100 \div 200$ nm and $a \sim 10 \div 20$ nm, we get for these two constants the following estimates: $\gamma_R \sim 10^{12}$ s$^{-1}$  and $|\Delta_L| \sim 10^{13}$ s$^{-1}$. One more parameter that should be chosen is the biexciton binding energy $\Delta_B$. The typical values of $\Delta_B$ are on the order of several meV, $\Delta_B \sim 2.5 \div 5~\mathrm{meV} \sim 10^{12}$~s$^{-1}$, although for some 2D systems, like transition metal dichalcogenides,~\cite{Mak2016,Mai2014} it can be one order of magnitude larger. In what follows, the spontaneous emission rate $\gamma$ is used as a unit of all frequency-dimensional quantities, whereas $\gamma^{-1}$ for time. According to our estimates, we set in $\gamma_R = 100\gamma$ and $|\Delta_L| = 1000\gamma$. The biexciton binding energy $\hbar\Delta_B$ is considered as a variable parameter.

The system of equations~(\ref{rho11})-~(\ref{Omega21}) represents a class of so-called stiff equations, characterized by several significantly different time scales. In our case, they are defined by $\gamma^{-1}\gg \gamma_R^{-1} \gg |\Delta_L|^{-1}$. It is well known that, under such conditions, general methods of numerical integration (such as the Runge-Kutta one) may lead to unreliable solutions. Bearing this in mind, we use special routines adapted for solving such kind of equations, in particular, the ODE23tb of MATLAB and some implementations of methods based on the backward differentiation formulas. 
%


%

\subsection{Steady-state analysis}
\label{steady-state}
As a first step in studying the supercrystal's optical response, we turn to the steady-state regime, setting to zero the time derivatives in in Eqs.~(\ref{rho11})~-~(\ref{Omega21}). Then we use the method developed in Appendix B to find the steady-state solutions to these equations. The results for different values of the biexciton binding energy $\hbar\Delta_B$ are presented in series of figures below.
\begin{figure}[ht!]
\begin{center}
\includegraphics[width=\columnwidth]{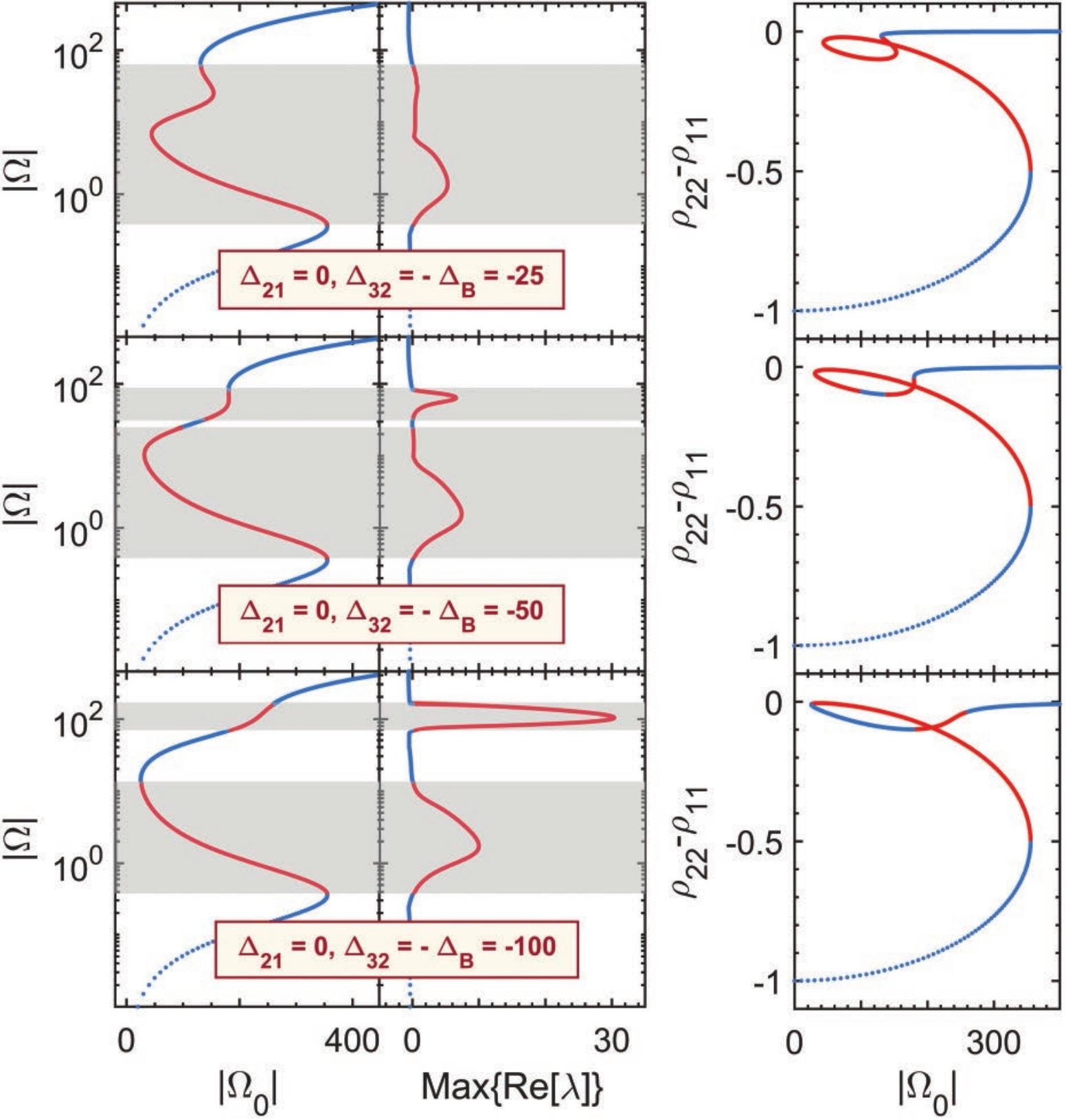}
\caption{\label{fig:one-photon_steady-state}  Steady-state solutions to Eqs.~(\ref{rho11})~-~(\ref{Omega21}) for the case of one-photon resonance ($\Delta_{21} = 0, \Delta_{32} = -\Delta_B$), obtained by the method described in Appendix B, for different values of the bi-exciton splitting $\Delta_B$ (shown in the plots). The leftmost column displays the $|\Omega|$-vs-$|\Omega_0|$ dependences, while the rightmost column - the $\rho_{22} - \rho_{11}$-vs-$|\Omega_0|$ dependences. The shaded regions show unstable parts of the stationary solutions, obtained by analyzing the Lyapunov exponents $\lambda$, the maximum values of the real parts of which, $\mathrm{Max}\{\mathrm{Re}[\lambda]\}$, are depicted in the middle panels.}
\end{center}
\end{figure}
\begin{figure}[ht!]
\begin{center}
\includegraphics[width=\columnwidth]{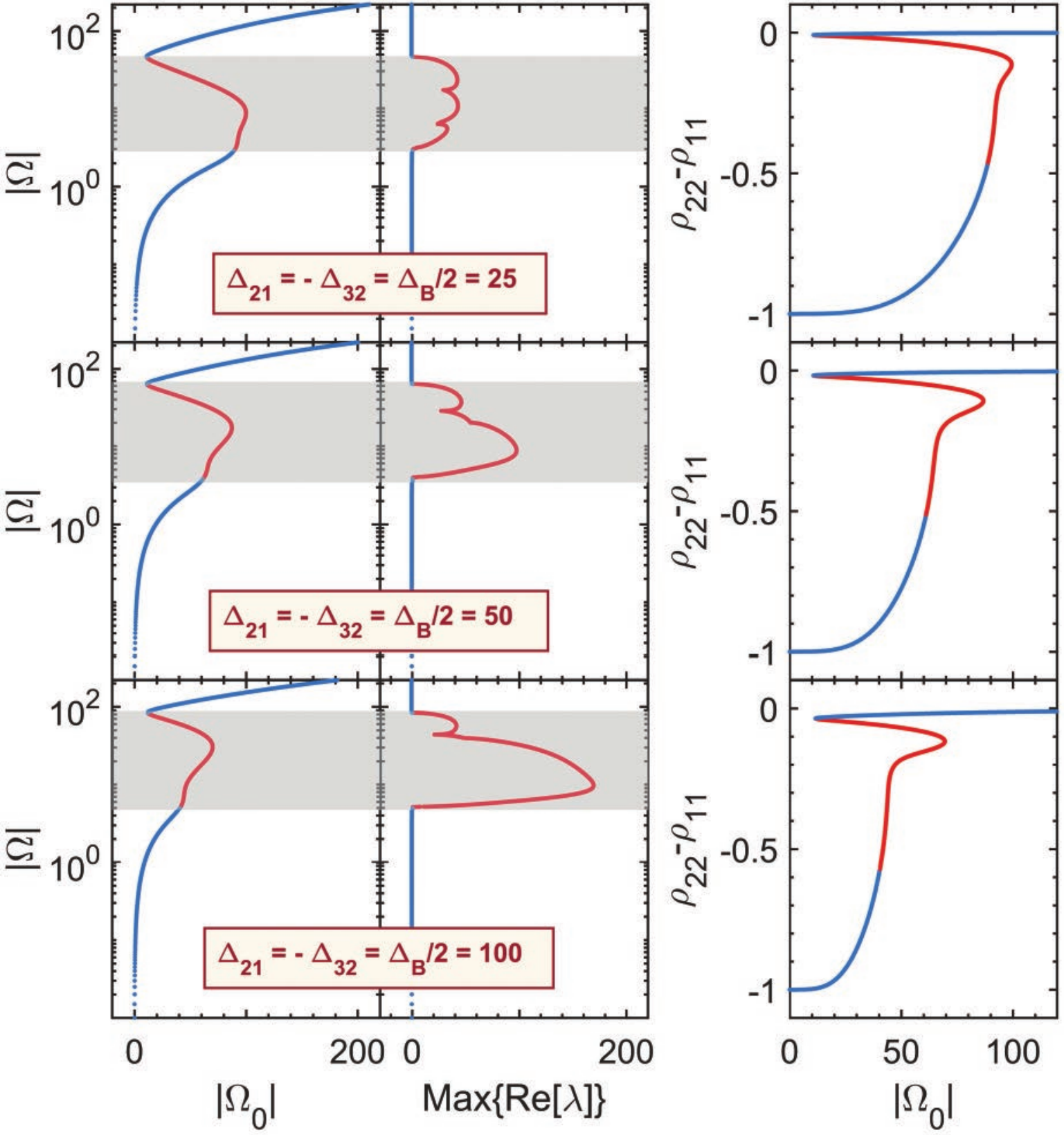}
\caption{\label{fig:two-photon_steady-state}  Same as in Fig.~\ref{fig:one-photon_steady-state}, only for the two-photon resonance ($\Delta_{21} = -\Delta_{32} = \Delta_B/2$).}
\end{center}
\end{figure}

Figures~\ref{fig:one-photon_steady-state} and~\ref{fig:two-photon_steady-state} show the dependence of the mean-field magnitude $|\Omega|$ (leftmost column) and the population difference $Z_{21} = \rho_{22} - \rho_{11}$ (rightmost column) on the external field magnitude $|\Omega_0|$ calculated for the one-photon ($\omega_0 = \omega_2,\, \Delta_{21} = 0,\, \Delta_{32} = -\Delta_B$) and two-photon ($\omega_0 = \omega_3/2,\, \Delta_{21} = -\Delta_{32} = \Delta_B/2$ resonance, respectively. As is seen from the figures, the mean-field magnitude $|\Omega|$ may have several solutions (up to five at $\Delta_{32} = -50$) for a given value of the external field magnitude $|\Omega_0|$, that signals emerging instabilities. We analyzed the stability of different branches, linearizing Eqs.~(\ref{rho11})~-~(\ref{Omega21}) around the steady-state solution and computing the Lyapunov exponents $\lambda$.~\cite{Eckmann1985,Katok1997} For this purpose, we calculated the eigenvalues of the Jacoby matrix of the linearized Eqs.~(\ref{rho11})~-~(\ref{Omega21}) as a function of $|\Omega|$. In our case, the Jacoby matrix is of the rank eight. Accordingly, there are eight complex-valued solutions for $\lambda$ for each value of $|\Omega|$. We selected from those an exponent with the maximal real part, $\mathrm{Max}\{\mathrm{Re}[\lambda]\}$, which determines the character of evolution of a small deviation from the steady-state solution. At $\mathrm{Max}\{\mathrm{Re}[\lambda]\} < 0$, a given solution is stable (deviation decreases) and {\it vice versa}. Calculated in this way $\mathrm{Max}\{\mathrm{Re}[\lambda]\}$ are plotted in the middle panels of Figs.~\ref{fig:one-photon_steady-state} and~\ref{fig:two-photon_steady-state}. The shaded regions show the unstable $(\mathrm{Max}\{\mathrm{Re}[\lambda]\} > 0$) parts of the steady-state solutions. We stress that not only branches with the negative slope are unstable, that is always the case, but those with the positive slopes as well. This occurs for both the one- and two-photon resonance conditions. Remarkably, for the case of the one-photon resonance, the upper branch of the steady-state solution only in part is unstable (see the plot for $\Delta_B = 100$). Moreover, for $\Delta_B = 50$, two unstable regions are separated by a stable one. The nature of the instabilities is analyzed in the next section.


\subsection{Time domain analysis}
\label{Time-domain}
To uncover the nature of instabilities of different branches of the steady-state solutions (shown in Figs.~\ref{fig:one-photon_steady-state} and~\ref{fig:two-photon_steady-state}), we solved numerically Eqs.~(\ref{rho11})~-~(\ref{Omega21}) for two conditions of excitation: (i) - tuning the external field into the one-photon resonance ($\omega = \omega_2$) and (i) - into the two-photon resonance ($\omega = \omega_3/2$). Two types of initial conditions were used: first, when the system initially is in the ground state ($\rho_{11} = 1$, while all other density matrix elements are equal to zero), and second, when the system starts from one of the steady-states solutions. We will present results for the most relevant points of the unstable branches, discussing the others only briefly.

\subsubsection{One-photon resonance ($\Delta_{21} = 0$, $\Delta_{32} = -\Delta_B$)}
\label{One-photon resonance}
\begin{figure}[ht!]
\begin{center}
\includegraphics[width=\columnwidth]{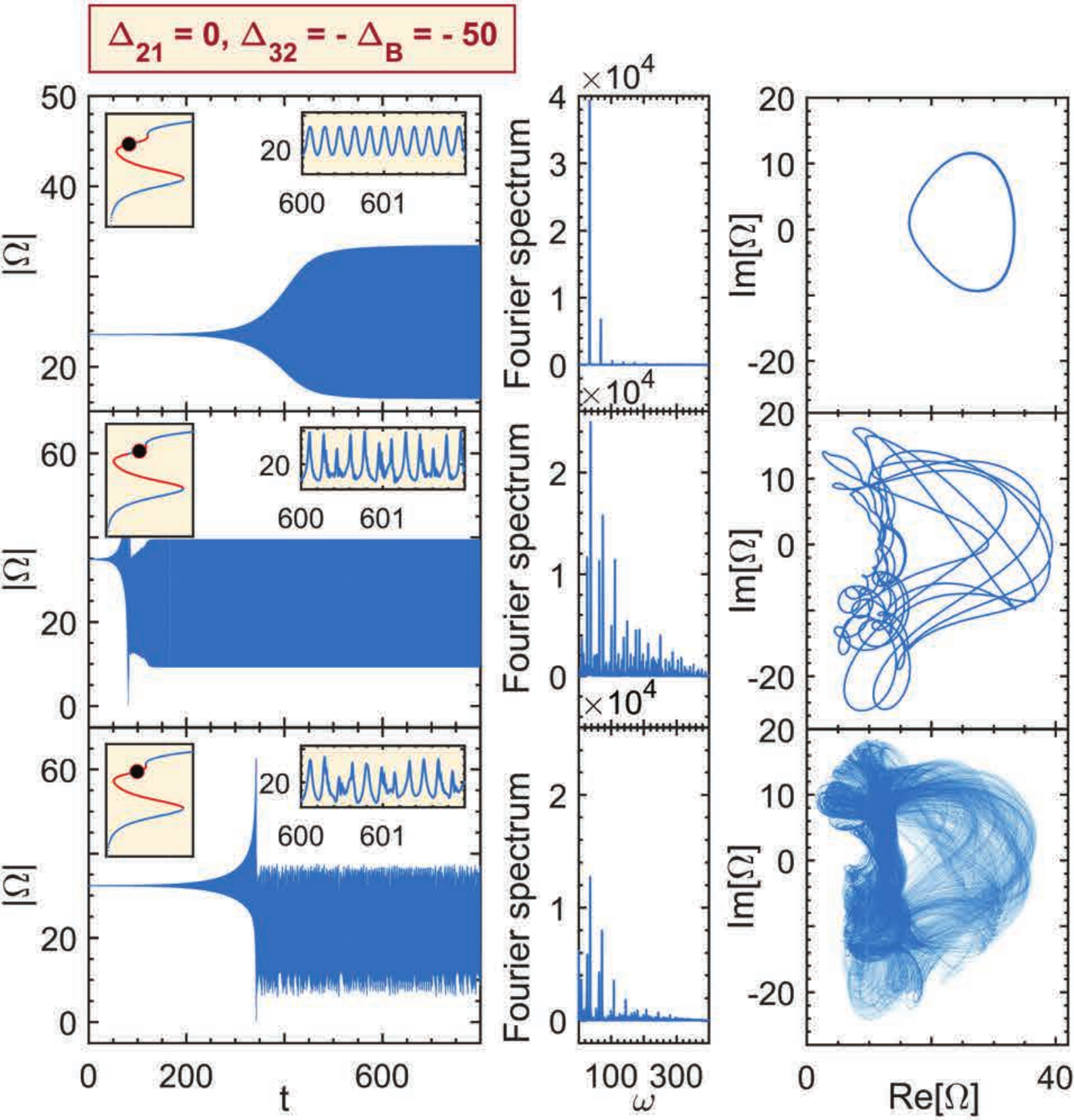}
\end{center}
\caption{\label{fig:DeltaB = 50} Time-domain dynamics of the mean-field magnitude $|\Omega|$ (left panels), the Fourier spectrum of the sustained signal (middle panels), and  phase space map of the sustained mean field $\Omega$ in the ($\mathrm{Re}[\Omega], \mathrm{Im}[\Omega]$) plane (right panels) for the case of the one-photon resonance ($\Delta_{21} = 0$, $\Delta_{32} = - \Delta_B$), obtained by solving Eqs.~(\ref{rho11})~-~(\ref{Omega21}) for $\Delta_B = 50$. The inserts show  fragments of the dynamics on a short time interval (upper-right insets) and the steady-state solution (upper-left insets), where the thick dots indicate the points on the steady-state solutions: $(|\Omega_0| = 90, |\Omega| = 23.5655)$ - upper row, $(|\Omega_0| = 150, |\Omega| = 34.7519)$  - middle row, and $(|\Omega_0| = 140.0002, |\Omega| = 32.3261)$ - lower row, for which the calculations were performed.}
\end{figure}
Figure~\ref{fig:DeltaB = 50} shows the results of time-domain calculations performed for the case of the one-photon resonance ($\Delta_{21} = 0$) setting $\Delta_{32} = -\Delta_B = -50$.  Three unstable points of the the steady-state solution were considered: ($|\Omega_0| = 90, |\Omega| = 23.5655$) - upper row, ($|\Omega_0| = 150, |\Omega| = 34.7519$) - middle row, and ($|\Omega_0| = 140.0002, |\Omega| = 32.3261$) - lower row. These points are indicated by the thick dots in the insets in the upper-left corners of the corresponding panels. Note that all these states of the system are inaccessible from the ground state: $\rho_{11}(0) = 1$, while all other density matrix elements are equal to zero. They can be accessed if one starts from a stable state at higher external field magnitude $|\Omega_0|$, further sweeps $|\Omega_0|$ down adiabatically until reaching the necessary point, and fixes here $|\Omega_0|$. Therefore, we consider that initially the system stands in the steady-state corresponding to the aforementioned points of the steady-state solution, and from this moment onward address its dynamics.

The leftmost column of panels of Fig.~\ref{fig:DeltaB = 50} shows the time evolution of the mean-field magnitude $|\Omega|$. As is seen, after some delay, the instabilities develop, the character of which significantly differs for all three considered points. The upper row demonstrates results for ($|\Omega_0| = 90, |\Omega| = 23.5655$); the dynamics looks like simple self-oscillations (see the upper-right inset for a blow up of the dynamics of $|\Omega(t)|$). The Fourier spectrum of the sustained signal (middle column) contains a couple of well-defined harmonics: see sharp peaks in the middle panel.
Accordingly, the phase map in the reduced phase space ($\mathrm{Re}[\Omega], \mathrm{Im}[\Omega]$) (right panel) represents a simple closed curve, commonly called limit cycle.~\cite{Eckmann1985,Katok1997}

For ($|\Omega_0| = 150, |\Omega| = 34.7519$), the dynamics is more complicated, nevertheless, revealing periodicity (see the inset in the left panel). The Fourier spectrum of the sustained signal is more complicated as compared to the previous case. It consists of a set of equidistant well-defined sharp peaks grouped in bands. Analyzing the relationship between different frequencies, we found that the their ratios appear to be rational numbers. Accordingly, the phase trajectory (right panel) represents a complicated but closed curve.

Contrary to the first two cases, for ($|\Omega_0| = 140.0002, |\Omega| = 32.3261$), the dynamics manifests signatures of aperiodic oscillations. This is reflected in the Fourier spectrum of the sustained signal, which also has a discrete nature but the peaks are smeared or have satellites and the frequency ratios are irrational, which is a signature of aperiodic oscillations.~\cite{Eckmann1985} The phase map of the sustained signal represents an open trajectory, almost filling a finite area in the phase space.

In all the above examples, the dynamics exhibits a delay before instabilities develop, which correlates excellently with the values of Lyapunov exponents for the corresponding points of the steady-state solution (see Fig.~\ref{fig:one-photon_steady-state}, the middle panel for $\Delta_B = 50$).
\begin{figure}[ht!]
\begin{center}
\includegraphics[width=\columnwidth]{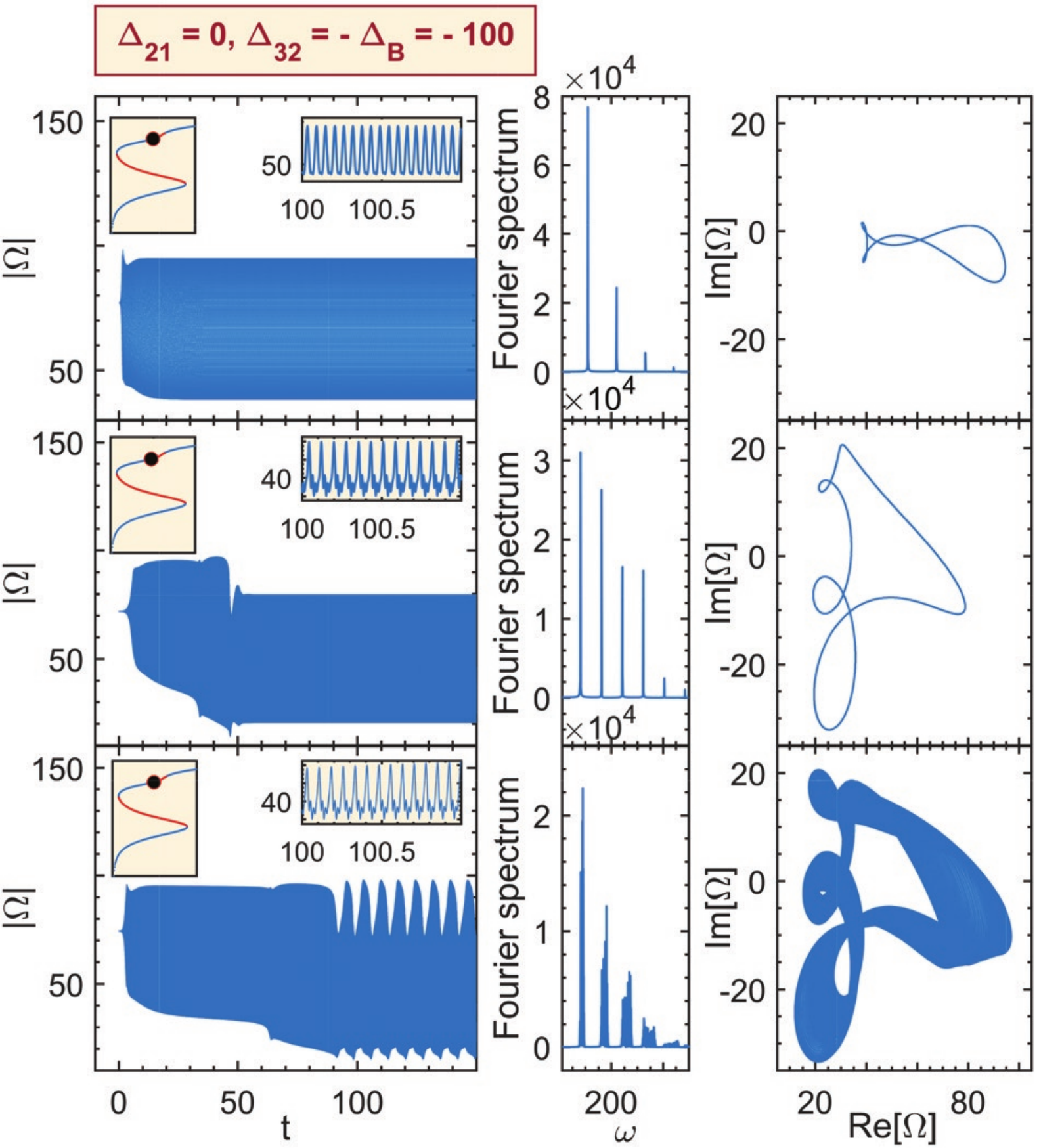}
\end{center}
\caption{\label{fig:DeltaB = 100} Same as in Fig.~\ref{fig:DeltaB = 50}, only calculated for $\Delta_B = 100$. The inserts in the upper-left corners show the steady-state solution, where the dots indicate the points $(|\Omega_0| = 200, |\Omega| = 76.9407)$  - upper row, $(|\Omega_0| = 190, |\Omega| = 71.9098)$ - middle row, and $(|\Omega_0| = 195, |\Omega| = 74.3081)$ - lower row, for which the calculations were performed.}
\end{figure}

In Fig.~\ref{fig:DeltaB = 100}, we present the results of the time-domain analysis of the system's dynamics for another value of the exciton binding energy $\Delta_B = 100$ ($\Delta_{32} = -100$). The calculations were performed for three points of the steady-state solution: $(|\Omega_0| = 200, |\Omega| = 76.9407)$ - upper row, $(|\Omega_0| = 190, |\Omega| = 71.9098)$ - middle row], and $(|\Omega_0| = 195, |\Omega| = 74.3081)$ - lower row.
As before, the left row of panels of Fig.~\ref{fig:DeltaB = 100} show the time evolution of the mean-field magnitude $|\Omega|$. The dynamics here essentially similar to the previous case of $\Delta_B = 50$, except the fact that the delay in developing the instabilities is short, which again correlates nicely with large positive values of the Lyapunov's exponents at the selected points of the steady-state solution (see Fig.~\ref{fig:one-photon_steady-state}, the middle panel for $\Delta_B = 100$ ).

Figure~\ref{fig:one-photon_steady-state} (middle panel) shows the maximum Lyapunov exponent as a function of $|\Omega|$. Also, not without interest is the whole set of eight Lyapunov's exponents. In Table~\ref{Table1}, we present it for $\Delta_B = 100$ and two particular  points of the steady-state solution, ($|\Omega_{0}| = 190,  |\Omega| = 71.9098$) and ($|\Omega_{0}| = 195,  |\Omega| = 74.3081$), for which the system's dynamics was calculated (see Fig.~\ref{fig:DeltaB = 100}). As follows from Table~\ref{Table1}, two complex conjugated Lyapunov's exponents have positive sign of their real parts. Generally, this can be a signature of the chaotic behavior of a nonlinear dynamical system, but not necessarily so,~\cite{Barrio2015} that is just our case: for the parameters indicated above, the system's dynamics demonstrates self- or aperiodic oscillations.

\begin{longtable}[c]{| c | c |}
 \caption{Lyapunov's exponents spectrum, $\Delta_B = 100$\label{Table1}}\\
 \hline
 \, $|\Omega_{0}| = 190,  |\Omega| = 71.9098$ \, & \, $|\Omega_{0}| = 195,  |\Omega| = 74.3081$ \,\\
 \hline
 \endfirsthead

 $1.14 + 137.76i$ & $2.41 + 145.50i$\\
 $1.14 - 137.76i$ & $2.41 - 145.50i$\\
 $-0.44 + 0.24i$ & $- 0.43 +	0.24i$\\
 $-0.44 - 0.24i$ & $-0.43 -	0.24i$\\
 $-7.66 + 183.86i$ & $-10.08 + 183.80i$\\
 $-7.66 - 183.86i$ & $-10.08 - 183.80i$\\
 $-13.46 + 266.72i$ & $-11.41 + 262.18i$\\
 $-13.46 - 266.72i$ & $-11.41 - 262.18i$\\
\hline
 \end{longtable}

\subsubsection{Two-photon resonance ($\Delta_{21} = -\Delta_{32} = -\Delta_B/2$)}
\label{Two-photon resonance}
\begin{figure}[ht!]
\begin{center}
\includegraphics[width=\columnwidth]{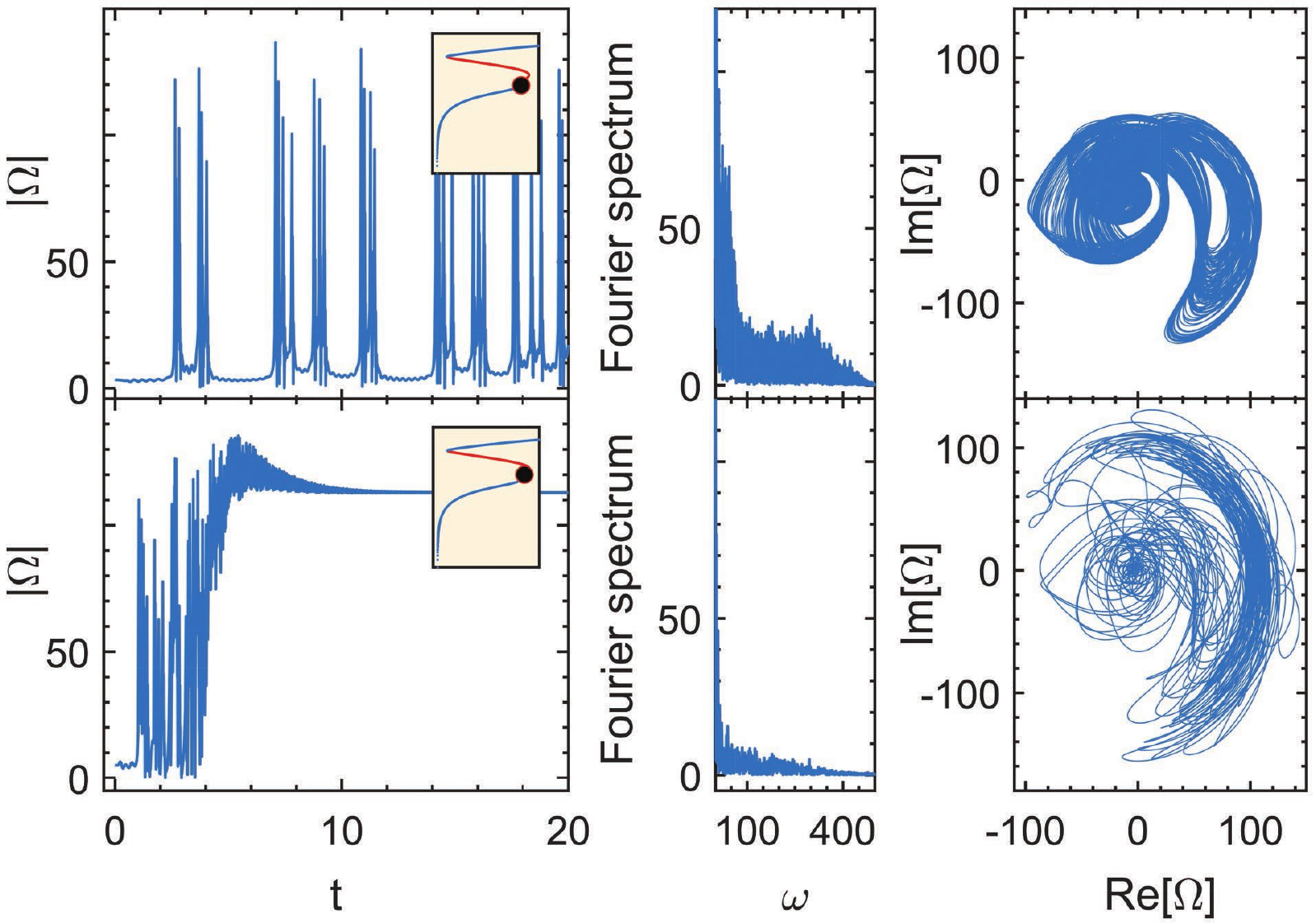}
\end{center}
\caption{\label{fig:Delta21 = -Delta32 = -25a} Time-domain dynamics of the mean-field magnitude $|\Omega|$ (left panels), its Fourier spectrum (middle panels), and  phase map of the full field $\Omega$ in the ($\mathrm{Re}[\Omega], \mathrm{Im}[\Omega]$) plane (right panels) for the case of the two-photon resonance ($\Delta_{21} = -\Delta_{32} = -\Delta_B/2$), obtained by solving Eqs.~(\ref{rho11})~-~(\ref{Omega21}) at $\Delta_{21} = -\Delta_{32} = - 25$ with the steady-state solution as the initial condition. The inserts show the points on the steady-state solutions (indicated by thick dots), $(|\Omega_0| = 91, |\Omega| = 23,5655)$ - upper row and $(|\Omega_0| = 94, |\Omega| = 32.3261)$ - lower row, for which the calculations were performed.}
\end{figure}

\begin{figure}[ht!]
\begin{center}
\includegraphics[width=\columnwidth]{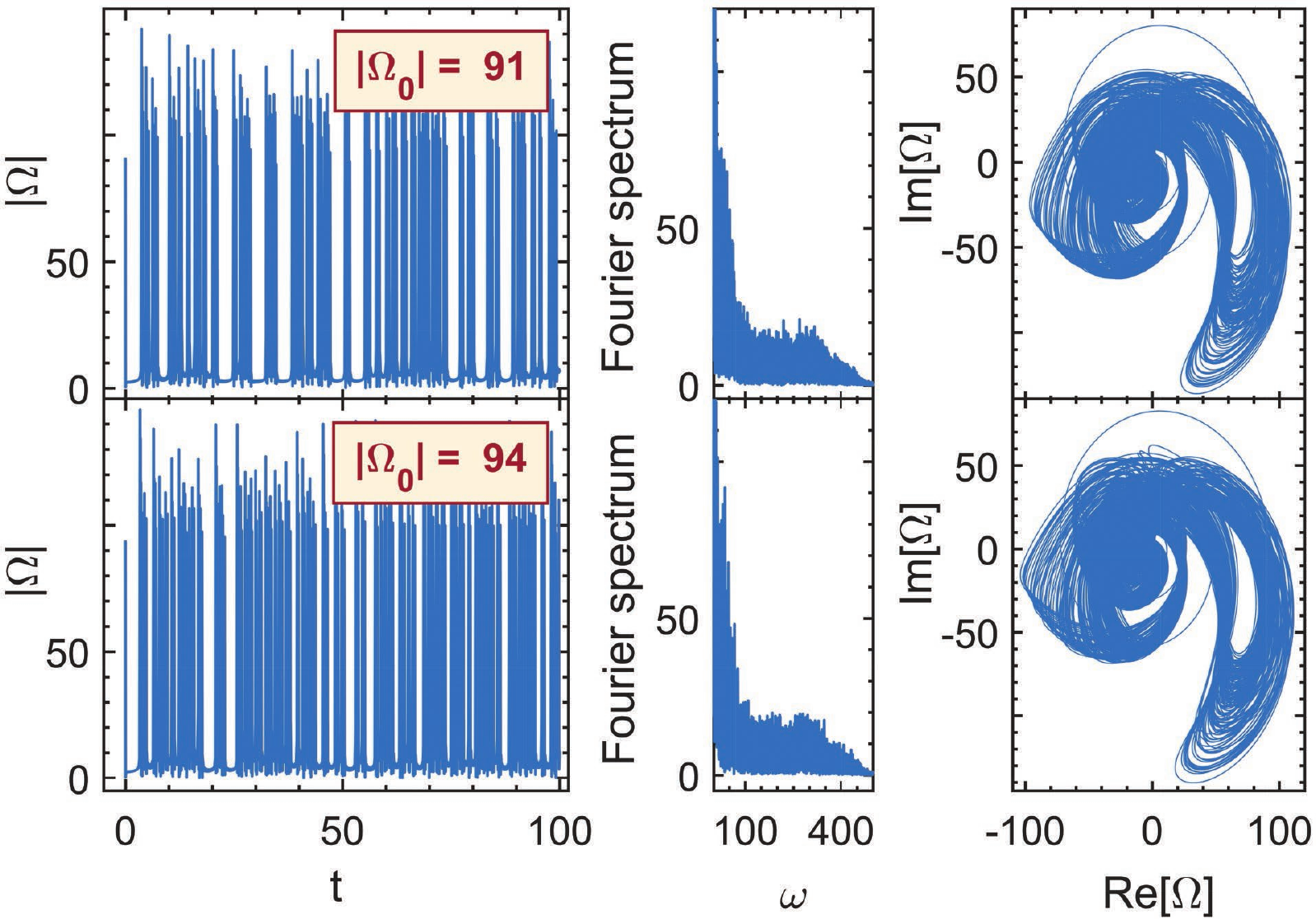}
\end{center}
\caption{\label{fig:Delta21 = -Delta32 = -25b} Same as in Fig.~\ref{fig:Delta21 = -Delta32 = -25a}, only calculated for the ground state as the initial condition and the same magnitudes of the external field as in Fig.~\ref{fig:Delta21 = -Delta32 = -25a}, $|\Omega_0| = 91$ (upper raw) and $|\Omega_0| = 94$ (lower raw).}
\end{figure}

In the case of the two-photon resonance ($\omega_0 = \omega_3/2$), a part of the lower branch of the steady state solution with a positive slope is unstable (see Fig.~\ref{fig:two-photon_steady-state}). In contrast with the previous case (see the preceding section), it is accessible from the system's ground state: $\rho_{11}(0) = 1$. Therefore, we will consider both options of the initial conditions: (i) the steady-state and (ii) the ground state. In Figs.~\ref{fig:Delta21 = -Delta32 = -25a} and~\ref{fig:Delta21 = -Delta32 = -25b}, the results of numerical calculations for these two cases are presented.

Figure~\ref{fig:Delta21 = -Delta32 = -25a} shows the time-domain calculations performed for $\Delta_B = 50$ using the steady-state solution as the initial condition. 
Two unstable points of the the steady-state solution were considered: ($|\Omega_0| = 91, |\Omega| = 23,5655$) - upper row, and $(|\Omega_0| = 94, |\Omega| = 32.3261)$ - lower row. These points are indicated by the thick dots on the steady-state solutions shown in the inserts.

We observe that for the point ($|\Omega_0| = 91, |\Omega| = 23,5655$), the dynamics demonstrates a highly irregular behavior. Note that in spite of   the signal is shown only on the time interval $t < 20$, such a behavior holds for much longer times. The Fourier spectrum of the signal represents a
broad structureless quasi-continuum, suggesting that the oscillatory regime found is of chaotic nature. The phase-space map confirms this, showing a trajectory that covers a finite volume of the phase space, not forming any close loop.

More insight on the character of the system's dynamics provides the Lyapunov's exponents spectrum shown in Table~\ref{Table2}. Remarcably, for the case under consideration, two different positive Lyapunov's exponents appear in the spectrum, that is a signature of emerging hyperchaos,~\cite{Barrio2015} which is in accordance with our numerical calculations.

\begin{longtable}[c]{| c | c |}
 \caption{Lyapunov's exponents spectrum, $\Delta_B = 50$\label{Table2}}\\
 \hline
 \, $|\Omega_{0}| = 91,  |\Omega| = 3.2816$ \, & \, $|\Omega_{0}| = 94,  |\Omega| = 5.0661$ \, \\
 \hline
 \endfirsthead

 10.98 & 31.62\\
 2.13 & 5.31\\
 $-0.09$ & $-0.07$\\
 $-2.24 + 0.80i$ & $-1.34$\\
 $-2.24 - 0.80i$ & $-6.81$\\
 $-12.19$ & $-13.41 + 85.29i$\\
 $-32.39 + 283.97i$ & $-13.41 - 85.29i$\\
 $-32.39 - 283.97i$ & $-33.84$\\
\hline
 \end{longtable}
By contrast, for the point $(|\Omega_0| = 94, |\Omega| = 32.3261)$ - lower row in Fig.~\ref{fig:Delta21 = -Delta32 = -25a}, the system, after a short chaotic stage, evolves towards the upper stable branch of the steady-state solution.

Figure~\ref{fig:Delta21 = -Delta32 = -25b} shows the evolution of the system being initially in the ground state ($\rho_{11}(0) = 1$, while all other density matrix elements are equal to zero) for the same external field magnitudes as in Fig.~\ref{fig:Delta21 = -Delta32 = -25a}, $|\Omega_0| = 91$ (upper row) and $|\Omega_0| = 94$ (lower row). Here, in both cases we observe hyperchaotic behavior of the system.

It should be pointed out the inherent difference between two calculations performed at the same external field magnitude $\Omega_0 = 94$ for different initial conditions (compare the dynamics of $|\Omega|$ plotted on the lower panels of Figs.~\ref{fig:Delta21 = -Delta32 = -25a} and~~\ref{fig:Delta21 = -Delta32 = -25b}). When choosing the initial condition, corresponding to the point on the steady-state solution, the long-time dynamics of $|\Omega|$ approaches the upper stable branch of the steady-state curve, whereas if the system is initially in the ground state, its time evolution demonstrates a hyper-chaotic behavior. This example shows that the scenario of the system's evolution may strongly depend on the initial conditions, resulting in the trajectories well separated and nonintersecting in the full phase space.

\subsubsection{Optical hysteresis}
\label{Optical hysteresis}
\begin{figure}[ht!]
\begin{center}
\includegraphics[width=\columnwidth]{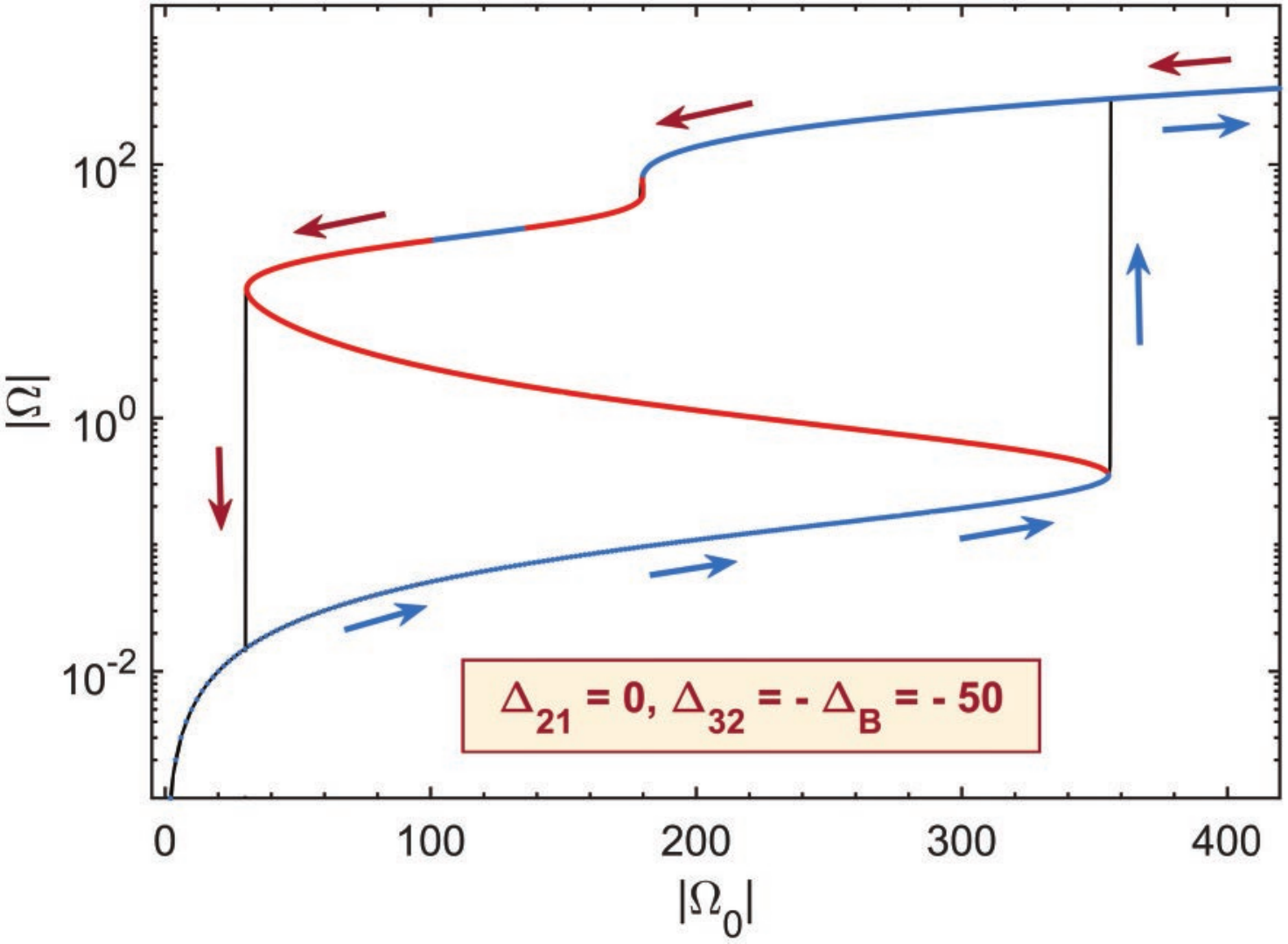}
\end{center}
\caption{\label{fig:Hysteresis one-photon DeltaB = 50} Optical hysteresis of the mean-field magnitude $|\Omega|$ calculated by solving Eqs.~(\ref{rho11})~-~(\ref{Omega21}) for the case of the one-photon resonance ($\Delta_{21} = 0$, $\Delta_{32} = - \Delta_B$) for $\Delta_B = 50$ under adiabatic scanning of the external field magnitude $|\Omega_0|$ up and down (shown by arrows).}
\end{figure}
%
\begin{figure}[ht!]
\begin{center}
\includegraphics[width=\columnwidth]{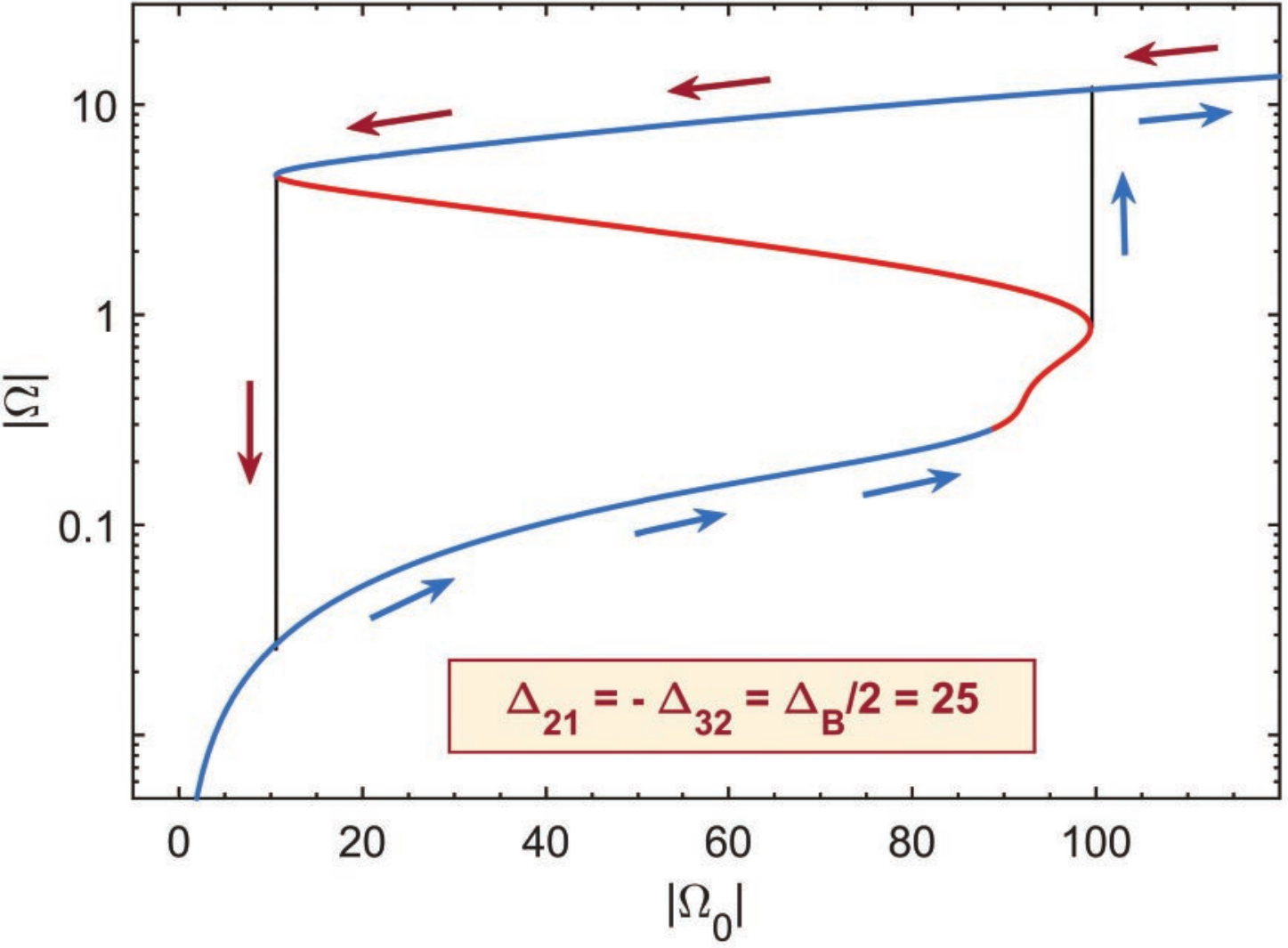}
\end{center}
\caption{\label{fig:Hysteresis two-photon DeltaB = 50} Same as in Fig.~\ref{fig:Hysteresis one-photon DeltaB = 50}, only calculated for the case of the two-photon resonance ($\Delta_{21} = -\Delta_{32} = -\Delta_B/2$) at $\Delta_B = 50$.}
\end{figure}

A multivalued steady-state response of the system, Figs.~\ref{fig:one-photon_steady-state} and~\ref{fig:two-photon_steady-state}, implies a hysteresis-like behavior of the latter under swiping the external field magnitude $|\Omega_0|$ adiabatically up and down. Figures~\ref{fig:Hysteresis one-photon DeltaB = 50} and~\ref{fig:Hysteresis two-photon DeltaB = 50} show the results for the one-photon and two-photon resonance, respectively.
In calculations, the external field magnitude $|\Omega_0|$  was swept linearly: $|\Omega_0| = 0.002 t$ for $t < T$) and $|\Omega_0| = 0.002 (t - T)$ for $t > T$ with the time step $\Delta t = 0.001$. From Figs.~\ref{fig:Hysteresis one-photon DeltaB = 50} and~\ref{fig:Hysteresis two-photon DeltaB = 50} it follows that in both cases, the supercrystal's optical response behaves in a bistable fashion. On swiping the external field magnitude $|\Omega_0|$ up, the system follows the lower branch of the steady state characteristics until it reaches the high-field turning point. Then it abruptly jumps up to the upper stable branch. Here, the system is saturated. On swiping $|\Omega_0|$ down, the system is remaining on the upper branch until $|\Omega_0|$ reaches the low-field turning point, where the system abruptly jumps down to the lower branch, forming finally the hysteresis loop. The negative-slope branches are not accessible in the adiabatic numerical experiment.

\subsection{Discussion}
\label{Discussion}
As was shown above, the system under consideration demonstrates a very rich optical dynamics: multistability, periodic and aperiodic aperiodic self-oscillations, and dynamical chaos. The origin of such a behavior is derived from the secondary field produced by the SQDs, which depends on the current state of SQDs. This can provide a strong enough positive feedback resulting finally in instabilities. On neglecting the secondary field, all above mentioned effects disappear.

Below, we discuss the principal nonlinearities responsible for the exotic SQD supercrystal optical response. For an illustration, let us consider Eqs.~(\ref{R21}) and~(\ref{R32}). Substituting  into Eqs.~(\ref{R21}) and~(\ref{R32}) the expression (\ref{Omega21}) for the field $\Omega$, one gets
\begin{widetext}
\begin{subequations}
\begin{eqnarray}
\label{R21 extended}
    \dot{\rho}_{21} = &-& \left[ i(\Delta_{21} - \Delta_L Z_{21}) + \frac{1}{2} - \gamma_R Z_{21} \right] \rho_{21}
    \nonumber \\
    &+& \mu(\gamma_R + i\Delta_L)Z_{21}\rho_{32}  + \mu(\gamma_R - i\Delta_L)(\rho_{21}^* + \mu\rho_{32}^*) \rho_{31}
    + \Omega_0 (Z_{21} + \mu\rho_{31})~,
\end{eqnarray}
\begin{eqnarray}
\label{R32 extended}
    \dot{\rho}_{32} = &-& \left[ i(\Delta_{32} - \mu^2\Delta_L Z_{32}) + \frac{1}{2} (1 + \mu^2) -\mu^2\gamma_R Z_{32} \right] \rho_{32}
    \nonumber \\
    &+& \mu(\gamma_R + i\Delta_L)Z_{32}\rho_{21} - (\gamma_R - i\Delta_L)(\rho_{21}^* + \mu\rho_{32}^*) \rho_{31}
    + \Omega_0 (\mu Z_{32} - \rho_{31})~,
\end{eqnarray}
\end{subequations}
\end{widetext}
As is seen, these equations contain a number of nonlinear terms, however a special attention should be paid to the first terms in the right-hand side, which describe the oscillations and decay of the off-diagonal density matrix element $\rho_{21}$ and $\rho_{32}$. It is found that the secondary field results in an additional frequency detuning $\Delta_L Z_{21}$ and damping $\gamma_R Z_{21}$ for $\rho_{21}$ and, respectively, $\mu^2\Delta_L Z_{32}$ and $\mu^2\gamma_R Z_{32}$ for $\rho_{32}$, depending on the corresponding population differences $Z_{21}$ and $Z_{32}$. Thus, a renormalization is evident: $\Delta_{21} \rightarrow \Delta_{21} - \Delta_L Z_{21}$ and $1/2 \rightarrow 1/2 - \gamma_R Z_{21}$ for the transition $2 \leftrightarrow 1$, and $\Delta_{32} \rightarrow \Delta_{32} - \mu^2\Delta_L Z_{32}$ and $1/2(1 + \mu^2) \rightarrow 1/2(1 + \mu^2) - \mu^2\gamma_R Z_{21}$ for the transition $3 \leftrightarrow 2$. Before the external field is switched on, the population difference $Z_{21} = -1$, whereas $Z_{32} = 0$, because the states $|2 \rangle$ and $|3 \rangle$ are unpopulated. Accordingly, only the lower ($1 \leftrightarrow 2$) transition experiences the above mentioned renormalization, whereas the higher ($2 \leftrightarrow 3$) transition does not. So the starting parameters are: the actual detuning away from the resonance $1 \leftrightarrow 2$ and the decay rate of the latter acquire values $\Delta_{21} - |\Delta_L|$ and $1/2 + \gamma_R$, respectively. As $|\Delta_L| \gg \Delta_{21}$ and $\gamma_R \gg 1/2$, namely $\Delta_L$ and $\gamma_R$ determine the resonance detuning and the decay rate of the $1 \leftrightarrow 2$ transition. All other resonance detunings and decay rates keep their bare values.

On switching the external field on, the system starts to evolve and react on the excitation in such a way that the secondary field, created by the system polarization and being in antiphase with the external field, almost compensates the latter. Indeed, let us consider the initial (linear) steady-state stage. Under this condition, the major contribution to the secondary field comes from $\rho_{21}$ which is given by
\begin {equation}
\label{eq:rho21 linear}
    \rho_{21} = - \frac{\Omega_0}{\frac{1}{2} + \gamma_R +i(\Delta_{21} + \Delta_L)}~.
\end{equation}
According to this, for the total field $\Omega$ one gets
\begin {equation}
\label{eq:Omega linear}
    \Omega =  \frac{\frac{1}{2} + i\Delta_{21}}{\frac{1}{2} + \gamma_R +i(\Delta_{21} + \Delta_L)}\Omega_0~.
\end{equation}
From Eq.~(\ref{eq:Omega linear}), it follows that for values of $\Delta_{21} \le 100$, $\gamma_R = 100$, and $|\Delta_L| = 1000$ used in our calculations, the inequality $|\Omega| \ll |\Omega_0|$ is always fulfilled, i.e., the secondary field destructively interferes with the incident field.

This (linear) stage holds until the acting field magnitude $|\Omega|$ becomes comparable with or larger than unity, $|\Omega| \gtrsim 1$. From this onwards, the system enters the strong excitation regime.  Accordingly, the dynamic shifts $\Delta_L Z_{21}$ starts to increase, whereas $\mu^2\Delta_L Z_{32}$ to decrease, improving the initially off-resonance situation towards a better resonance with both transitions. As a result, a redistribution of the level populations and the competition between transitions come into play, which finally creates conditions for emerging instabilities (see for more details Ref.~\onlinecite{Nugroho2017}).

The bistability and hysteresis loop are ensured by the parameters $\Delta_L$, $\gamma_R$, and $\Delta_B$ which are chosen in the calculations. They are well above the bistability threshold.~\cite{Friedberg1989,Malyshev2012}

\section{Reflectance}
\label{Reflectance}
In our analysis of the layer's nonlinear response we used the total field $\Omega$ acting on an emitter. In an experiment, the reflected or transmitted fields are commonly detected. These two fields  differ from the total field: they are determined by the fare-zone part of $\Omega$ and are given by the following expressions:
\begin{subequations}
\begin{equation}
\label{Reflected field}
\Omega_\mathrm{refl} = \gamma_R (\rho_{21} + \mu\rho_{32})~.
\end{equation}
\begin{equation}
\label{Transmitted field}
\Omega_\mathrm{tr} = \Omega_0 + \gamma_R (\rho_{21} + \mu\rho_{32})~.
\end{equation}
\end{subequations}
The reflectance and transmittance, $R$ and $T$, respectively, are then defined as
\begin{equation}
\label{Reflection and Transmission}
R = \left|\frac{\Omega_\mathrm{refl}}{\Omega_0}\right|^2, \quad T = \left|\frac{\Omega_\mathrm{tr}}{\Omega_0}\right|^2~.
\end{equation}

Let us look first at the linear regime of excitation and restrict ourselves to analyzing reflectance. Substituting Eq.~(\ref{eq:rho21 linear}) into Eq.~(\ref{Reflection and Transmission}), for the reflectance $R$ we get
\begin{equation}
\label{Reflectance linear}
    R = \left| \frac{\gamma_R}{\frac{1}{2} + \gamma_R +i(\Delta_{21} + \Delta_L)}\right|^2~.
\end{equation}
From this expression it follows that for the detuning $\Delta_{21} \le 100$ used in our calculations, the reflectance $R \ll 1$, because $|\Delta_L| \gg \Delta_{21}, \gamma_R$. Nevertheless, all features of the mean field $\Omega$ found in our study will be mirrored in the reflected field $\Omega_\mathrm{refl}$ and the reflectance as well.

Remarkably, if we are in the vicinity of the resonance renormalized by the near field, {\it i. e.}, $\Delta_{21} \approx -\Delta_L$, the reflectance of the system is close to unity, $R \approx 1$. Thus, in this region of frequencies, the SQD supercrystal appears to be {\it a perfect reflector}. Recently, it has been reported that on the basis of a monolayer of MoSe$_2$, an atomically thin mirror can be realized.~\cite{Back2018,Scuri2018} Our considered supercristall represents one more example of such a system but it offers various ways to control the properties of the mirror and the working frequency range by geometry and material choice.

\begin{figure}[ht!]
\begin{center}
\includegraphics[width=\columnwidth]{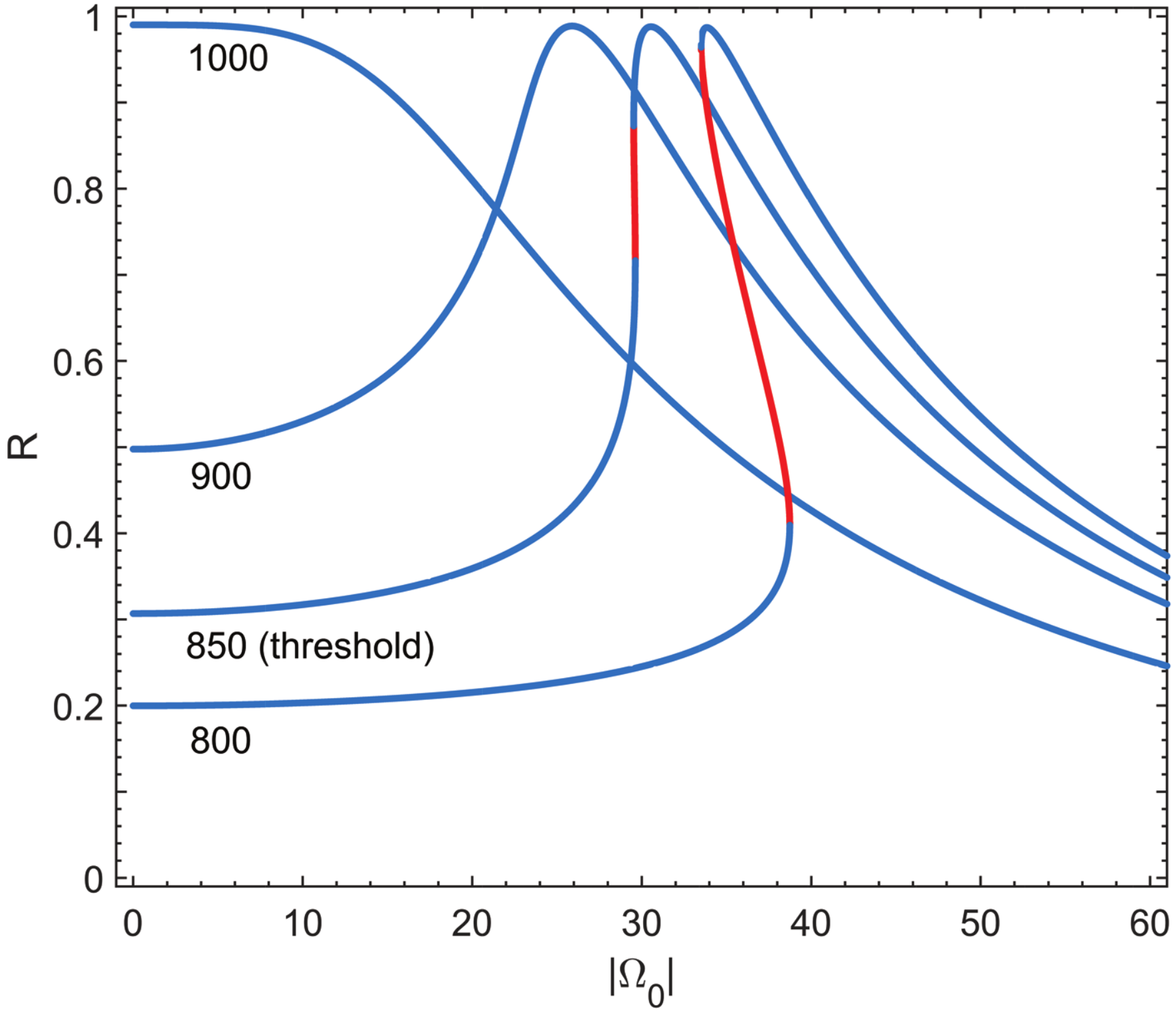}
\end{center}
\caption{\label{fig:Reflection DeltaB = 50} The steady-state reflectance $R$, Eq.~(\ref{Reflection and Transmission}), versus the detuning $\Delta_{21}$ in the vicinity of the renormalized by the near field resonance, $\Delta_{21} \leq - \Delta_L$, calculated  for the biexciton binding energy $\Delta_B = 50$. The values of $\Delta_{21}$ considered are shown in the plot, herewith $\Delta^{th}_{21} = 850$ being the threshold for bistability to occur. The red fragments of the curves indicate the unstable branches.}
\end{figure}

It is of great interest to look at the nonlinear behaviour of the reflectance in the vicinity of $\Delta_{21} = -\Delta_L$. We performed calculations of the $\Delta_{21}$-dependence of the reflectance $R$ in the range of $\Delta_{21} \leq -\Delta_L$ (above the renormalized resonance). The results almost do not depend on the biexciton binding energy $\Delta_B$, so for the illustration purpose, we chosen $\Delta_B = 50$. Figure~\ref{fig:Reflection DeltaB = 50} shows the corresponding result.

From Fig.~\ref{fig:Reflection DeltaB = 50} one observes that at the exact resonance ($\Delta_{21} = -\Delta_L$), the reflectance monotonously decreases (from almost unity) on increasing the external field magnitude $|\Omega_0|$. This behavior has its natural explanation in the population difference dependence of the actual detuning $\Delta_{21}^\prime = \Delta_{21} - \Delta_L Z_{21}$ [see Eq.(\ref{R21 extended})]: on excitation, the system goes away from the renormalized resonance ($\Delta_{21} = -\Delta_L, \Delta_{21}^\prime = 0$) and the reflectance reduces accordingly.

By contrast, if the system is far from the renormalized resonance (above, $\Delta_{21} \leq -\Delta_L$), the low-field reflectance drops down and perfectly follows Eq.~(\ref{Reflectance linear}). Increasing the applied field magnitude $|\Omega_0|$ and subsequent excitation of the system improves the resonance conditions as $\Delta_{21} - \Delta_LZ_{21} \to 0$, that manifests itself in high reflectance, $R \approx 1$. Furthermore, starting some critical value of $\Delta_{21}$, namely $\Delta^{th}_{21} = 850$ for the set of parameters used, the reflectance becomes three-valued, indicating the existence of the optical bistability. The critical value $\Delta^{th}_{21} = 850$ is in a good agreement with the theoretical estimate within a framework of an effective two-level model, $\Delta_{21} = -\Delta_L - \sqrt{3}\gamma_R \approx 827$.~\cite{Benedict1990} A small deviation from the calculated value originates from the fact that the system under consideration has an additional biexciton level, a small admixture of which slightly affects the threshold value.

\section{Summary and outlook}
\label{Summary}
We conducted a theoretical study of the optical response of a two-dimensional semiconductor quantum dot supercrystal subject to a single-frequency quasiresonant excitation. An isolated SQD was modeled as a three-level ladder-like system with the ground, one-exciton and biexciton states. The set of parameters used in our study is typical for SQDs emitting in the visible range, such as, for instance, CdSe and CdSe/ZnSe.
We took into account the SQD-SQD interaction within the framework of the mean-field approximation. An exact method of solving the nonlinear steady-state problem, developed in the paper, allowed us to reveal the fact that the system's response can be multivalued. Analyzing the Lyapunov exponents, we found windows of stability and instability of different branches of the steady-state solutions. It turned out that the supercrystal optical response might demonstrate bistability, self-oscillation, and dynamic chaos/hyperchaos under a single CW excitation.~\cite{footnote}
The frequency of self-oscillations depends on the external field magnitude and, for the set of parameters used, falls in the THz region.
We have provided a physical insight into the instabilities found, which have their origin in the competition between the ground-to-one exciton and one exciton-to-biexciton  transitions, driven by the near-field SQD-SQD interactions.

On the basis of our findings, a SQD supercrystal can be viewed as (i) an all-optical bistable switch, (II) as a tunable generator of trains of THz pulses (in self-oscillation regime), and (iii) as a noise generator (in chaotic regime). In addition, the sensitivity of the supercrystal's optical response to the initial conditions, experienced in a chaotic/hyperchaotic regime, is of interest for encryption of information.~\cite{Gao2008} And finally, the 2D SQD supercrystal may act as a bistable nanosized mirror. All this makes such systems promising objects for practical applications in all-optical information processing and optical computing.

One problem which remains to be solved is the validity of the mean-field approximation for a sample of finite size. As a mater of fact, the surrounding of SQDs at the boundary differs from that inside the sample. Because of that, this inhomogeneity will propagate through the sample due to the SQD-SQD dipole-dipole interaction. A question arises: To what extent the finite size effects will violate the results of the mean-field (homogeneous) approximation? The same problem holds if only a part of the sample is subjected to irradiation. These issues are beyond the scope of the present paper and will be the objectives of a forthcoming study.

\acknowledgments
P. A. Z. and A. V. M. acknowledge support from Spanish MINECO grants MAT2013-46308 and MAT2016-75955.
I. V. R. acknowledges support from the Russian Foundation for Basic Research, project no. 15-02-08369.

\begin{appendix}
\section{Numerical evaluation of $\gamma_R$ and $\Delta_L$}
\label{ApendixA}
\begin{figure}[ht!]
\begin{center}
\includegraphics[width=\columnwidth]{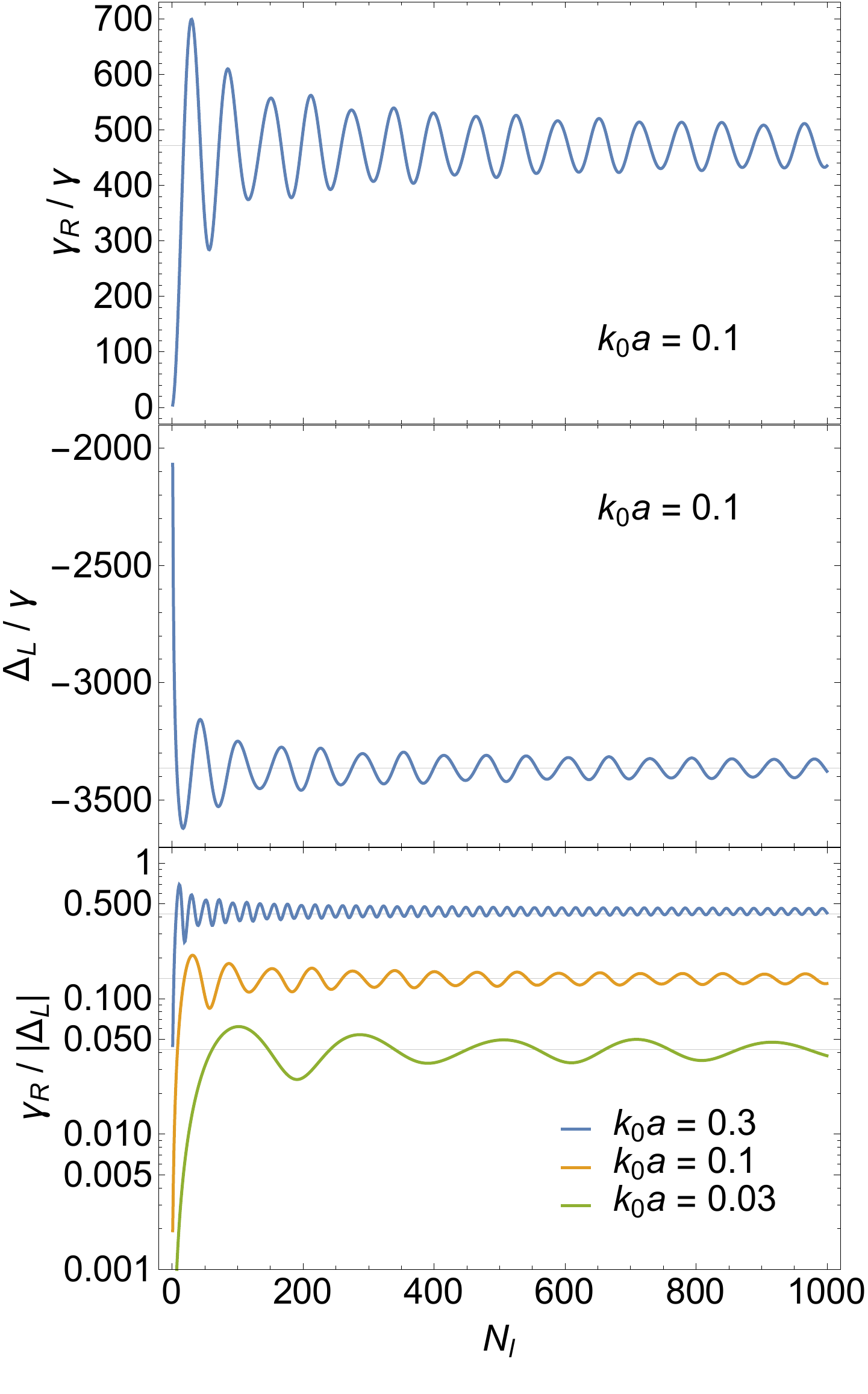}
\caption{\label{fig:gammaR and DeltaL} The lateral size dependence of the collective radiation rate $\gamma_R$ (upper plot), the near-zone dipole-dipole interaction of SQDs, $\Delta_L$ (middle plot), and the ratio $\gamma_R/|\Delta_L|$ calculated from Eq.~(\ref{gammaR}) and Eq.~(\ref{DeltaL}) for different values of $k_0a$ (indicated in the plots). Thin horizontal lines are guides for the eye, showing the mean around which the oscillations of the underlined quantities occur.}
\end{center}
\end{figure}
Here, we evaluate numerically $\gamma_R$ and $\Delta_L$, given by Eqs.~(\ref{gammaR}) and~(\ref{DeltaL}), for a large system ($N_xa, N_ya \gg \lambdabar$), setting in $N_x = N_y = N_l$. In Fig.~\ref{fig:gammaR and DeltaL}, we plotted $\gamma_R$, $\Delta_L$, and $\gamma_R/|\Delta_L|$ against the lattice lateral size $N_l$ for different values of $k_0a$. As is seen, all these quantities reveal small oscillations around averaged values, which originate from a bad convergency of the sums, containing terms proportional to $|{\bf n}|^{-1}$. Analyzing these data under the assumption of the $(k_0a)^{-3}$-scaling of $\Delta_L$, we derived that the numerical coefficient in the averaged value of $\Delta_L$ is $\approx -3.35$. From the ratios $\gamma_R/\Delta_L$, it then follows the $(k_0a)^{-2}$-scaling of the averaged value of $\gamma_R$ with the numerical factor $\approx 4.51$. These scalings describe excellently all numerical data presented in Fig.~\ref{fig:gammaR and DeltaL}, leading finally to Eqs.~(\ref{gammaRextended}) and~(\ref{DeltaLextended}).

\section{Solution of the steady-state problem}
\label{AppendixB}
The steady-state problem is governed by the following set of equations:
%
%
\begin{subequations}
\begin{equation}
\label{a:rho22}
\rho_{22} + \Omega \rho_{21}^* + \Omega^* \rho_{21} = 0~,
\end{equation}
\begin{equation}
\label{a:rho33}
\mu \rho_{33} + \Omega \rho_{32}^* + \Omega^* \rho_{32} = 0~,
\end{equation}
\begin{equation}
\label{a:R21}
\Omega(\rho_{22} - \rho_{11}) - \left( i\Delta_{21} + \frac{1}{2} \right) \rho_{21}  + \mu \Omega^* \rho_{31} = 0~,
\end{equation}
\begin{equation}
\label{a:R32}
\mu \Omega(\rho_{33} - \rho_{22}) - \left[ i\Delta_{32} + \frac{1}{2} (1 + \mu^2) \right] \rho_{32}  - \Omega^* \rho_{31} = 0~,
\end{equation}
\begin{equation}
\label{a:R31}
- \left( i\Delta_{31} + \frac{1}{2}\mu^2 \right) \rho_{31} - \mu \Omega \rho_{21} + \Omega \rho_{32} = 0~,
\end{equation}
\begin{equation}
\label{a:Normalization}
\rho_{11} + \rho_{22} + \rho_{33} = 1~.
\end{equation}
\end{subequations}
%
%
Thus, originally the system of nine nonlinear coupled equations for the density matrix elements should be solved to find the dependence of these elements and the full field $\Omega$ on the external field $\Omega_0$. The two fields are related by Eq.~(\ref{Omega21}) which we rewrite here for convenience in the following form:
\begin{equation}
\label{Omega0}
\Omega_0 = \Omega - (\gamma_R + i\Delta_L)(\rho_{21} + \mu\rho_{32})\ .
\end{equation}

Traditionally, one or another numerical method of direct solution of the nonlinear system (\ref{a:rho22}-\ref{a:Normalization}) is used. Below we propose a much more efficient and essentially linear parametric method to solve this nonlinear problem.

First, we note that Eqs.~(\ref{a:rho22}-\ref{a:Normalization}) and Eq.~(\ref{Omega0}) are invariant under the following phase transformation:
\begin{subequations}
\begin{equation}
\rho_{21} \mapsto \rho_{21}\,e^{i\,\varphi},\;\;
\rho_{31} \mapsto \rho_{31}\,e^{i\,\varphi},\;\;
\rho_{32} \mapsto \rho_{32}\,e^{2i\,\varphi}
\label{rhophasetrans}
\end{equation}
\begin{equation}
\Omega \mapsto \Omega\,e^{i\,\varphi},\quad
\Omega_0 \mapsto \Omega_0\,e^{i\,\varphi}\ ,
\label{Omegaphasetrans}
\end{equation}
\label{phasetrans}
\end{subequations}
where $\varphi$ is an arbitrary phase. 

Second, Eq.~(\ref{Omegaphasetrans}) suggests that instead of (naturally) considering the external field $\Omega_0$ to be real, one can consider the full field $\Omega$ to be real. In this case, given that $\Omega^\star=\Omega$, the system of Eqs.~(\ref{a:rho22}-\ref{a:Normalization}) are linear in the density matrix elements if $\Omega$ is considered to be a real parameter. Hence, the system can be solved and the {\it unique} parametric dependence of the density matrix elements on $\Omega$ can be obtained. Further, Eq.~(\ref{Omega0}) provides the {\it unique} parametric dependence of the external field $\Omega_0$ on the real full field $\Omega$.

Finally, the sought dependencies of the density matrix elements on the external field $\Omega_0$ are obtained in the parametric way, varying the real $\Omega$ within an appropriate interval of values. To recover the ``traditional'' case, in which the external field $\Omega_0$ is real, the transformations (\ref{phasetrans}) can be used with the phase $\varphi$ given by:
\begin{equation}
\varphi = -\mathrm{arg}\left[
\, \Omega - (\gamma_R + i\Delta_L)(\rho_{21} + \mu\rho_{32}) \,
\right]\ .
\end{equation}

To conclude, we note that our proposed method is general for a whole class of the steady-state Maxwell-Bloch-like equations and can be applied for a broad range of systems.

\end{appendix}

\bibliography{sqd-array}
\bibliographystyle{apsrev4-1}

\end{document}